\documentclass[prx, reprint, nofootinbib, 10pt, showkeys]{revtex4-2} 
\usepackage{etex, babel}

\usepackage{amsmath, amssymb, braket, commath, bm, empheq, array}

\usepackage[]{graphicx}
\graphicspath{{images/}}
\DeclareGraphicsExtensions{.eps,.jpg,.png,.pdf}
\usepackage{hyperref}
\hypersetup{
	colorlinks=true,
	linkcolor=red,
	filecolor=blue,
	urlcolor=magenta,
}

\usepackage{dblfloatfix}

\usepackage{booktabs, multirow}
\newcommand{\aloc}{\ensuremath{\alpha_\textrm{loc}}}		
\newcommand{\bte}{\ensuremath{\beta\textrm{-ensemble}}}
\def\idel{\ensuremath\delta\textrm{I}}		

\def\eg{\ensuremath E_{\textrm{G}}}			
\def\cale{\ensuremath \mathcal{E}}			
\def\eloc{\ensuremath E_\textrm{loc}}		
\newcommand{\ferf}[1]{\ensuremath{\text{Erf}\del{#1}}}

\newcommand{\fgamma}[1]{\ensuremath{\Gamma\del{#1}}}
\newcommand{\fdg}[1]{\ensuremath{\text{digamma}\del{#1}}}

\newcommand{\get}{\ensuremath{\gamma_{\text{ET}}}}	
\newcommand{\ipr}{\ensuremath{\textrm{I}}}

\newcommand{\mean}[1]{\ensuremath{\left\langle#1\right\rangle}}

\newcommand{\nloc}{\ensuremath{N_\textrm{loc}}}		
\newcommand{\prob}[1]{\ensuremath{\text{P}\del{#1}}}
\newcommand{\shn}{\ensuremath{\textrm{S}}}
\newcommand{\sloc}{\ensuremath{ \ket{\Psi_\textrm{loc}} }}
\newcommand{\zl}{\ensuremath{\zeta_l}}
\begin{document}
\title{Absence of Mobility Edge in Short-range Uncorrelated Disordered Model: \\ Coexistence of Localized and Extended States}
\author{Adway Kumar Das$^1$}\email{akd19rs062@iiserkol.ac.in}
\author{Anandamohan Ghosh$^1$}\email{anandamohan@iiserkol.ac.in}
\affiliation{
	$^1$Indian Institute of Science Education and Research Kolkata, Mohanpur, 741246 India
}
\author{Ivan M. Khaymovich$^{2,3}$}\email{ivan.khaymovich@gmail.com}
\affiliation{$^2$Nordita, Stockholm University and KTH Royal Institute of Technology Hannes Alfv\'ens v\"ag 12, SE-106 91 Stockholm, Sweden}
\affiliation{$^3$Institute for Physics of Microstructures, Russian Academy of Sciences, 603950 Nizhny Novgorod, GSP-105, Russia}

\date{\today}
\begin{abstract}
Unlike the well-known Mott's argument that extended and localized states should not coexist at the same energy in a generic random potential, we provide an example of
a nearest-neighbor tight-binding disordered model which carries both localized and extended states without forming the mobility edge (ME).
Unexpectedly, this example appears to be given by a well-studied \bte\ with independently distributed random diagonal potential and inhomogeneous kinetic hopping terms.
In order to analytically tackle the problem, we locally map the above model to the 1D Anderson model with matrix-size- and position-dependent hopping and confirm the coexistence
of localized and extended states, which is shown to be robust to the perturbations of both potential and kinetic terms due to the separation of the above states in space.
In addition, the mapping shows that the extended states are non-ergodic and allows to analytically estimate their fractal dimensions.
\end{abstract}
\maketitle
A mobility edge~(ME)~\cite{Mott1967}, separating localized and extended states in disordered systems, has been established and studied for decades.
Known to be present in various semiconductors, amorphous media and even in disordered liquid metals, ME has become a hallmark of the Anderson~\cite{Anderson1} and many-body~\cite{Luitz1} localization transitions.
It is commonly believed that in any short-range model, with random uncorrelated entries, just below the localization transition ME separates the localized and extended states in the energy spectrum. Therefore, eigenstates with different localization properties cannot coexist at the same energy for the same system parameter values. The argument behind this, given by Nevill F. Mott~\cite{Mott1967}, is straightforward: if extended and localized states coexist at the same energy, any perturbation of the disorder potential immediately hybridizes them, making both extended. In this Letter, we provide an example of one-dimensional (1D) disordered short-range model, where in any realization both extended or localized states can emerge at a given energy. Hence, disorder averaging forbids the ME formation. In this work, we identify the necessary conditions to be satisfied to observe such coexistence.

First, the system should avoid level degeneracy or attraction, i.e.~it should possess some (residual) level repulsion. Indeed, any resonance in the energy levels, corresponding to localized and extended states, should be suppressed in order to observe their coexistence without hybridization. Among short-range uncorrelated models, the natural ensemble for tunable and controllable level repulsion is the so-called \bte, represented by real symmetric tridiagonal matrices, with independent random elements~\cite{Dumitriu1}. Such an ensemble is parameterized by the Dyson's index $\beta$ and has the same joint probability distribution of eigenvalues like in the well-known Gaussian random-matrix ensembles~\cite{Mehta1}, but for any real $\beta$
along with $\beta=1$, $2$, $4$.
The limit $\beta\to 0$ yields uncorrelated eigenvalues as observed in integrable systems~\cite{Berry3}, whereas $\beta\geq 1$ produces correlated spectra as in chaotic systems~\cite{Bohigas1}.

Second, to suppress the overlap of the localized and extended states, the latter ones should be non-ergodic, having a support set smaller than the entire Hilbert space.
It is known that the presence of disorder may break ergodicity in some quantum systems, while they remain nonintegrable, thus, delocalized~\cite{Luca1, Garcia2, Ray1, Wang3, Pino2, Pino3, Das4}.
Such non-ergodic extended (NEE) phase is associated with a non-trivial scaling of the eigenstate fluctuations~\cite{Fyodorov1} and can be captured by various random matrix models~\cite{Kravtsov1, Nosov2019correlation,Nosov2019mixtures,Khaymovich1, Cizeau1, Das3,LN-RP_RRG,LN-RP_dyn,Kutlin2021emergent,Tang2022nonergodic,Motamarri2021RDM}, hierarchical graphs~\cite{Tikhonov2}, and many-body disordered systems~\cite{Luitz1,Luitz3,Pino2}. Quite surprisingly, it has been recently shown that the above mentioned \bte\ also hosts the NEE phase over an extended parameter regime~\cite{Das2}.

\begin{figure}[h!]
	\centering
	\includegraphics[width=0.45\textwidth]{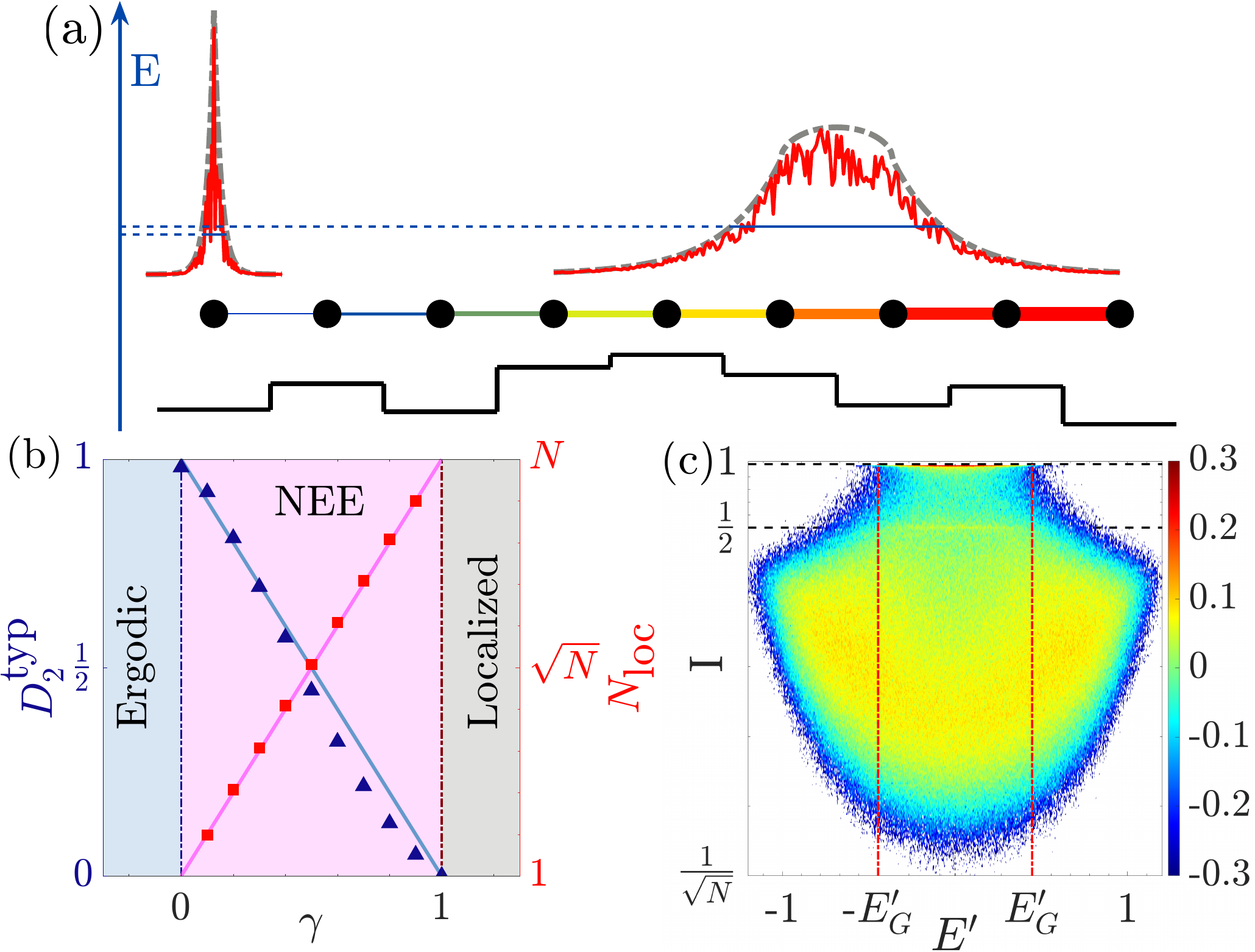}
	\caption{(a)~Schematic of \bte, given by 1D lattice in Eq.~\eqref{eq_H_elements}. The hopping increases along the lattice (thickness, color of links), yields localized (left) and extended (right) states, coexisting at the same energy, but living in spatially different system parts.
		(b)~Phase diagram of \bte, with three distinct phases along with typical fractal dimension in the NEE phase, $D_2^\text{typ}\approx 1-\gamma$, and the 
number $\nloc \sim N^\gamma$ of strongly localized states, $\sloc$.
		(c)~Joint density~$\prob{\ipr, E'}$ of IPR, $\ipr$ and the energy, $E' = E/\epsilon_\beta$, rescaled to the bandwidth $\epsilon_\beta$ for $\gamma = 0.7$. The colorbar indicates the values of joint density in $\log_N$ scale, $\del{-\eg', \eg'}$ is the rescaled
		energy band for coexistent states.
		Numerical results are for $N = 8192$ and $128$ realizations.
	}
	\label{fig_lattice}
\end{figure}

The last, but not the least, the NEE states should be separated in space from the localized ones.
The construction of \bte~\cite{Dumitriu1} introduces inhomogeneity, where the distributions of hopping matrix elements $y_n$ significantly depend on the lattice coordinate $n$, see Fig.~\ref{fig_lattice}(a) and Eq.~\eqref{eq_hop_typical}. Consequently the eigenstates of \bte\ become spatially separated.

In this Letter, we show that by fulfilling all of the above three crucial ingredients, \bte\ provides an ideal platform for realizing coexistent localized and extended eigenstates. We numerically confirm that $\mathcal{O}(\beta^{-1})$ localized states coexist along with the extended states within the middle of the spectral band without forming any ME. In addition, the NEE phase of \bte\ is shown to exhibit anomalies in the spectral statistics: nearby eigenvalues remain uncorrelated, while two distant eigenvalues, separated by $\Delta E>\del{N\beta}^{-\frac{1}{2}}$, can be correlated. Here and further, $N$ is the system size. Such a feature is in sharp distinction from the NEE states observed in a paradigmatic Rosenzweig-Porter ensemble (RPE)~\cite{Rosenzweig1,Kravtsov1, Khaymovich1}. This last aspect unveils the origin of the absence of the ME and can be analytically explained by a local equivalence of \bte\ to an 1D Anderson model with $N$-dependent hopping strength. This equivalence demonstrates that \bte\ separates into nearly independent blocks, where localized and (non-ergodic) extended eigenstates appear to be located in spatially separated blocks, but share nearly the same spectral energies.


\bte\ is composed of the matrices $H$, with the following nonzero elements~\cite{Dumitriu1}
\begin{align}
	\label{eq_H_elements}
	\begin{split}
		H_{n, n} &= x_n,\quad H_{n, n+1} = H_{n+1, n} = y_n\\
		x_n &\sim \mathcal{N}(0, 1),\quad \sqrt{2}y_n\sim \chi_{n\beta}
	\end{split}
\end{align}
where $\mathcal{N}(0, 1)$ is the normal distribution and $\chi_k$ is the chi-distribution with a degree of freedom $k$. $H$ represents a 1D lattice with an open boundary, where a particle can randomly hop to the nearest neighbors under disordered on-site potentials, Fig.~\ref{fig_lattice}(a). The relative strengths of on-site potentials $\cbr{x_n}$ and the hopping amplitudes $\cbr{y_n}$ make it convenient to re-parameterize $\beta$ as $\beta = N^{-\gamma}$, leading to the typical behavior of hopping amplitudes, see Appendix~\ref{sec_hop}
\begin{align}
	\label{eq_hop_typical}
	y_n^\text{typ} \sim \begin{cases}
		\exp\del{-\dfrac{N^\gamma}{n}}, &n< N^\gamma\\
		\sqrt{\dfrac{n}{N^\gamma}}, &n> N^\gamma
	\end{cases}.
\end{align}

Thus, on average $y_n$ increases across the lattice and presents a highly inhomogeneous system. Due to such inhomogeneity, \bte\ hosts three distinct phases: ergodic ($\gamma\leq 0$), NEE ($0<\gamma<1$) and localized ($\gamma\geq 1$), separated by second-order phase transitions~\cite{Das2}. In the localized phase, all the levels are uncorrelated and Poisson-distributed as the eigenstates are localized with a finite support in the thermodynamic limit ($N\to\infty$), see the left state in Fig.~\ref{fig_lattice}(a). Contrarily in the ergodic phase, all energies are correlated irrespective of their distance and the bulk eigenstates are extended over the entire Hilbert space. Different phases in \bte\ exist due to its inhomogeneous hopping terms, otherwise phase transition is absent in generic 1D systems with uncorrelated short-range hopping~\cite{Cuesta1, Albert1}.

The Hilbert-space structure in the NEE phase can be understood from the system-size scaling of the inverse participation ratio (IPR), $\ipr_j = \sum_{n = 1}^{N}|\Psi_j(n)|^4$ of the eigenstate at the energy $E_j$ having $n$th component $\Psi_j(n)$. It is observed that $\mean{\log \ipr_j} \simeq -D_2^\text{typ} \log N$, where $D_2^\text{typ} \approx 1-\gamma$ is a typical fractal dimension, quantifying the eigenstate support set, Fig.~\ref{fig_lattice}(b). Hence most of the eigenstates occupy an extensive number, but vanishing fraction of the Hilbert space in the NEE phase. However, the density of IPR shows a peak around $\ipr = 1$ indicating the presence of strongly localized states, $\sloc$, along with a finite fraction of extended states with $\ipr\ll 1$, see Appendix~\ref{sec_psi_loc}. 
We consider a small tolerance value $\idel\ll 1$ and identify $\sloc$ as a state with $\ipr>1-\idel$. In Fig.~\ref{fig_lattice}(b), we show that the ensemble-averaged number of strongly localized states, $\nloc\propto N^\gamma$ coexisting with the finite fraction of NEE states.

Now, looking at the joint density of IPR and energy, Fig.~\ref{fig_lattice}(c), we unveil the spectral structure of $\sloc$. In the NEE phase of \bte, $\sloc$ appears only within an energy window $(-\eg, \eg)$, centered around mid-spectrum ($E = 0$). The ensemble average $\mean{\eg}\propto N^{\alpha_G}$ is nearly $N$-independent, with $\alpha_G\ll 1$, irrespective of $\delta\ipr\ll 1$, see Appendix~\ref{sec_e_loc}. 
Importantly, Fig.~\ref{fig_mobility}(c) indicates that within $(-\eg, \eg)$, IPR takes a wide range of values from $\mathcal{O}(1)$ to $N^{-D_2}$, convincingly demonstrating the coexistence of localized and extended states.

\begin{figure}[t]
	\centering
	\includegraphics[width=0.4\textwidth]{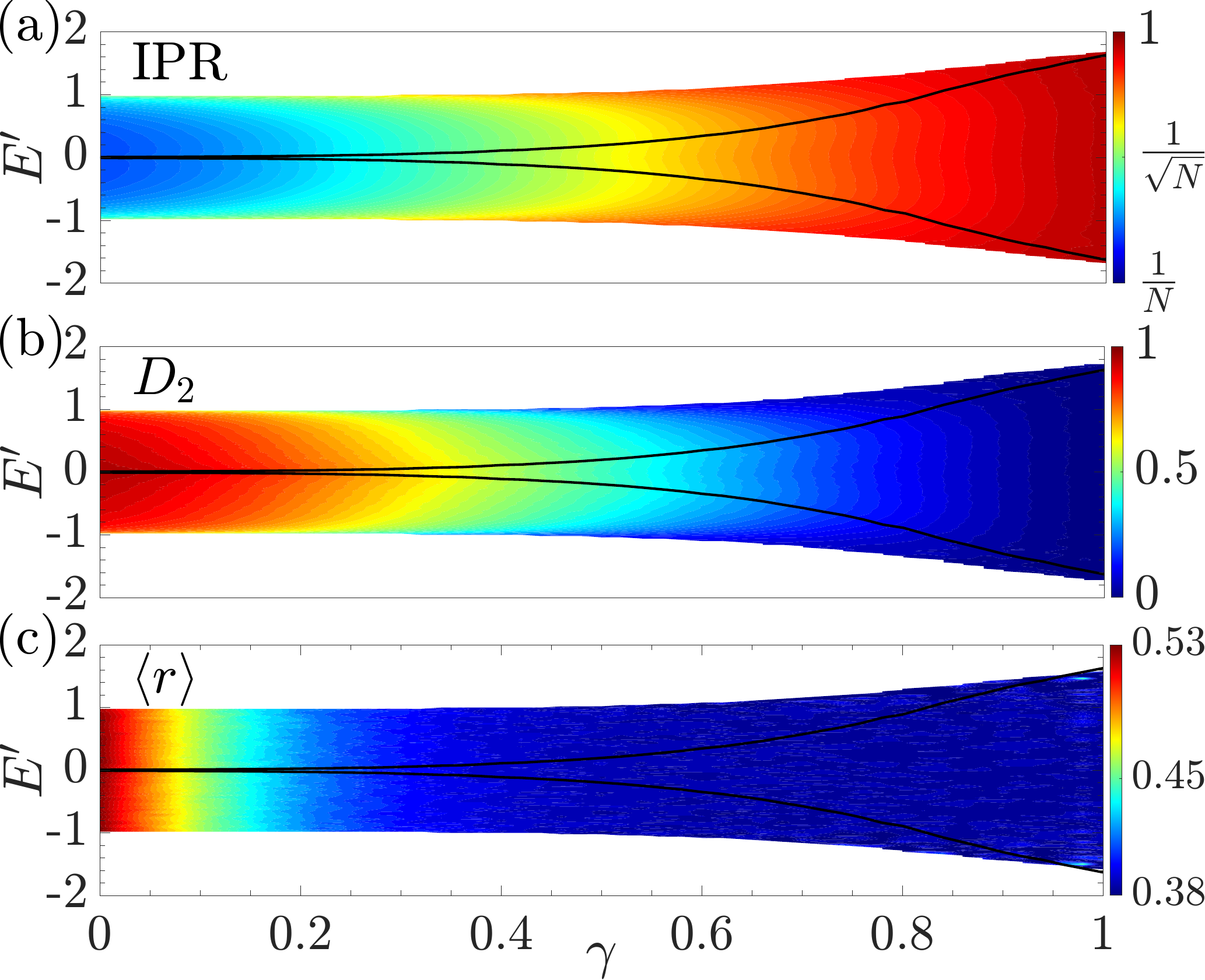}
	\caption{(a)~IPR, (b)~fractal dimension $D_2$, (c)~mean level-spacing ratio $r$ for $N = 8192$ in the $\gamma$-$E'$ plane. Solid black lines indicate $\del{-\eg', \eg'}$, the energy band of localized states.}
	\label{fig_mobility}
\end{figure}
Does this coexistence form a ME? ME has been observed in the L\'evy ensemble~\cite{Cizeau1, Metz3}, quasi-periodic lattice~\cite{Gopalakrishnan2,Deng2,Wang_mosaic,Wang_mosaic,DasSarma1,DasSarma2,DasSarma3, Wang3, Ahmed2022flat}, 3-D Anderson model~\cite{Bulka1}, quantum random energy model~\cite{Scardicchio_QREM} and many-body localization~\cite{Luitz1}. In order to search for ME in the \bte, first, we compute the energy-dependent IPR
\begin{align}
	\label{eq_I_e}
	\ipr(E) = \frac{1}{N\rho(E)}\sum_{j = 1}^{N} \ipr_j\delta\del{E-E_j} \ ,
\end{align}
where $\rho\del{E}$ is the global density of states (DOS).
For a given energy $\ipr(E)\to 0$ ($\simeq\mathcal{O}(1)$) for extended (localized) states. Hence, an existence of a ME would have implied $\ipr(E)$ exhibiting an energy-dependent crossover from $0$ to $\mathcal{O}(1)$ within $(-\eg, \eg)$. Fig.~\ref{fig_mobility}(a), showing IPR in the $\gamma$-$E'$ plane of the energy $E' = E/\epsilon_\beta$ , rescaled by the global energy bandwidth $\epsilon_\beta = 2\sqrt{\mean{E^2}} = \sqrt{4+2N^{1-\gamma}}$, provides no evidence of ME at any value of $\gamma$.

However, IPR has a fat-tailed distribution in the \bte\ and may not be a self-averaging quantity~\cite{Das2}, requiring more convincing measures. Thus, we extract the energy-dependent fractal dimension $D_2(E)$ from the system-size scaling of median$(\ipr)$ within small windows across the energy spectrum. Fig.~\ref{fig_mobility}(b) shows that $D_2(E)$ is energy independent in $(-\eg, \eg)$ irrespective of $\gamma$. This convincingly shows that the ME is absent in the NEE phase of \bte\ despite the coexistence of $\mathcal{O}(N^\gamma)$ localized states around $E = 0$.

\begin{figure}[t]
	\centering
	\includegraphics[width=0.4\textwidth]{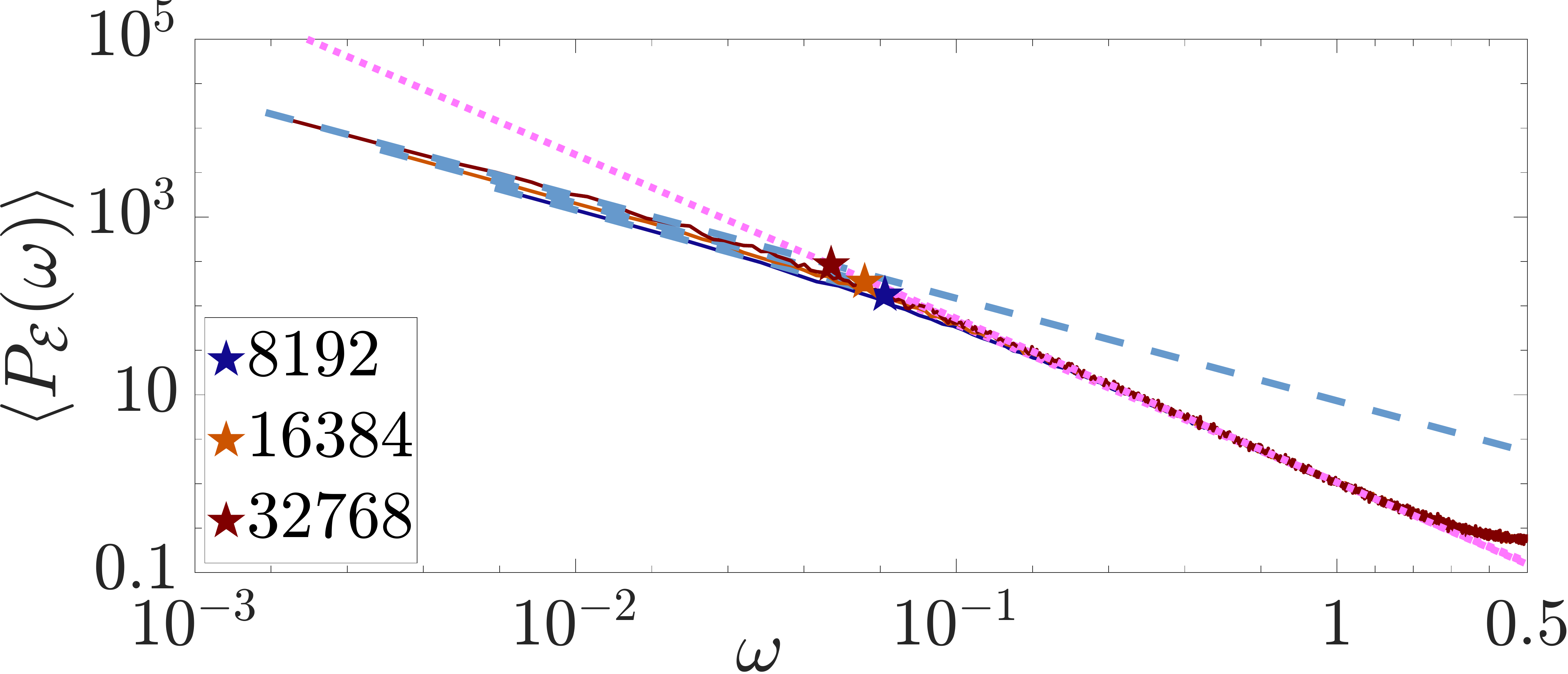}
	\caption{Ensemble-averaged power spectrum vs.~dimensionless frequency for $\gamma = 0.4$ in a randomly chosen energy window with width $\Delta\cale = \frac{N}{8}$. Dashed (dotted) lines show $\omega^{-1}$ ($\omega^{-2}$) fits and the star denotes the critical frequency.
	}
	\label{fig_power}
\end{figure}
In addition to the spectral homogeneity of the localization properties, one can consider energy-level correlations across the spectrum. The ensemble-averaged level-spacing ratio $\mean{r}$~\cite{Oganesyan2007,Atas1} exhibits criticality around $\gamma = 0$, implying that the neighboring eigenvalues are typically uncorrelated in the NEE phase~\cite{Das2} in thermodynamic limit, having some residual level repulsion $\beta=N^{-\gamma}$ at finite sizes. In order to understand the energy dependence of short-range energy correlation we compute
\begin{align}
	\label{eq_r_E}
	\begin{split}
		r(E) &= \frac{1}{N\rho\del{E}}\sum_{n = 1}^{N}r_n\delta\del{E - E_n}\\
		r_n &= \min\del{\tilde{r}_n, \frac{1}{\tilde{r}_n}},\: \tilde{r}_n = \frac{E_{n+1} - E_n}{E_n - E_{n-1}}
	\end{split}.
\end{align}
The ensemble-averaged $r$ in the $\gamma$-$E'$ plane, Fig.~\ref{fig_mobility}(c), also shows no energy-dependent crossover from Poisson, $\mean{r}\approx 0.38$, to Wigner-Dyson, $\mean{r}\approx 0.53$, statistics~\cite{Oganesyan2007,Atas1}, for all $\gamma$. Thus, short-range energy correlations are also homogeneous over the bulk energy spectrum.

Besides the spectral homogeneity at short energy scales, one needs to characterize long-range two-level correlation. This can be captured by the fluctuations of the $n$th unfolded energy level, $\cale_n$~\cite{Guhr1} around its mean position, $n$ via the power spectrum $P(\omega)$ vs frequency $\omega$, Fourier-dual to $n$.
For Poisson (Wigner-Dyson) statistics, the power spectrum of the fluctuations $\delta_n\equiv \cale_n - n$ decays as $\omega^{-2}$ ($\omega^{-1}$)~\cite{Faleiro1, Riser1,Relano3}.
In \bte, $P(\omega)$ shows a heterogeneous behavior in the frequency domain, see Fig.~\ref{fig_power}:
\begin{align}
	\label{eq_P_d_w_whole}
	P(\omega)\propto \begin{cases}
		\omega^{-1}, & \omega<\omega_c\\
		\omega^{-2}, & \omega>\omega_c
	\end{cases},\quad \omega_c = \frac{\pi}{N^\gamma}.
\end{align}
The critical frequency, $\omega_c$, corresponds to the unfolded energy scale $N^\gamma$ such that two unfolded energy levels $\cale_{1, 2}$ are correlated if $\abs{\cale_1 - \cale_2} > N^\gamma$, and uncorrelated otherwise. Therefore, in an energy window $\del{\cale - \frac{\Delta\cale}{2}, \cale + \frac{\Delta\cale}{2}}$, $P_\cale(\omega)\sim \omega^{-2}$ shows only Poisson behavior for $\Delta\cale<N^\gamma$ irrespective of $\cale$. Such a long-range correlation is unusual and complimentary to the energy correlations typically observed in various models like RPE~\cite{Kravtsov1, Berkovits2, Das1}, deformed Poisson ensemble~\cite{Das3}, or driven Aubry-Andr\'e models~\cite{Ray1}. Usually the eigenstates hybridize below the Thouless energy, $\cale_\text{Th}$~\cite{Altshuler3}, while distant eigenvalues, separated by $\Delta \cale>\cale_\text{Th}$, remain uncorrelated~\cite{Tomasi1,LN-RP_dyn}.

Above numerical results unambiguously show that the coexistence of localized and extended states fails to form any ME.
This can be analytically understood from a local equivalence of the \bte\ to an Anderson model. Indeed, in \bte, the hopping amplitudes over the first $\mathcal{O}(N^\gamma)$ sites are much smaller in magnitude in comparison to the typical on-site potentials $x_n^\text{typ}\sim \mathcal{O}(1)$. Hence the sites $1,2,\cdots, n\lesssim N^\gamma$ can be considered to be effectively disconnected from the rest of the lattice and hosts $\sloc$.
Corresponding DOS follows the normal distribution, while both short- and long-range energy correlations are given by the Poisson statistics~\cite{supple}.

The structure of the eigenstates at the remaining sites can be understood as follows.
The hopping amplitudes for lattice sites $n> N^\gamma$, given by Eq.~\eqref{eq_hop_typical}, are self-averaging and homogeneous
with small relative fluctuations $\abs{y_{n+\delta n}-y_{n}}\ll y_{n}$ for $\delta n\ll n$.
Consequently, one can partition the entire lattice into the spatial blocks $n \in \Delta_l$, where hopping is approximately the same and
\begin{align}
	\label{eq_part}
	\Delta_l\equiv [N^{\gamma+\zl}, N^{\gamma+\zeta_{l+1}}] ~~.
\end{align}
Here $\zl\equiv l\delta\zeta$ with $\delta\zeta\simeq \frac{O(1)}{\log N}$, and $0\leq l<\frac{1-\gamma}{\delta\zeta}$.
The number of sites in $\Delta_l$ is $N_l\approx N^{\gamma+\zl}$ and the hopping amplitudes
can be written as $y^\text{typ}_l \approx \sqrt{\frac{N^{\gamma+\zl}}{N^\gamma}} = N^{\frac{\zl}{2}}$. Thus within $\Delta_l$, hopping amplitudes are effectively constant and the model is equivalent to the 1D Anderson model with hopping strength $N^{\frac{\zl}{2}}$ and on-site potential $\mathcal{O}(1)$.
As a result, within each block, the eigenstates exponentially decay, $\abs{\Psi(j)}\sim \exp\del{- \frac{\abs{j - j_\text{loc}}}{\xi}}$, see Appendix~\ref{sec_exp}, 
with
the localization length $\xi_l\sim \frac{\mean{y_n^2}}{\mean{x_n^2}} = N^{\zl}$~\cite{Sanchez1}
and localization centers $j_\text{loc}$ randomly distributed in $\Delta_l$.
Furthermore, as each sub-block of length $\xi_l\sim N^{\zl}$ has $N^{\zl}$ eigenstates, thus, each block $\Delta_l$ can accommodate $N_l/\xi_l \approx N^{\gamma}$ sub-blocks.
The eigenstates within each sub-block hybridize, but not across sub-blocks.
The local Anderson equivalence implies that each sub-block has Gaussian DOS with a bandwidth~\cite{Furedi1}
\begin{align}
	\label{eq_block_DOS}
	\Delta E_l \simeq \sqrt{ \frac{1}{N^{\zl}} \sum_{i_l\in \Delta_l} N^{\zl}} = N^\frac{\zl}{2}
\end{align}
and the corresponding mean level spacing is $\delta_l = \frac{\Delta E_l}{N^{\zl}} = N^{-\frac{\zl}{2}}$.
The above equivalence
explains the properties of the NEE phase of \bte\ as illustrated below.

\begin{figure}[t!]
	\centering
	\includegraphics[width=0.38\textwidth]{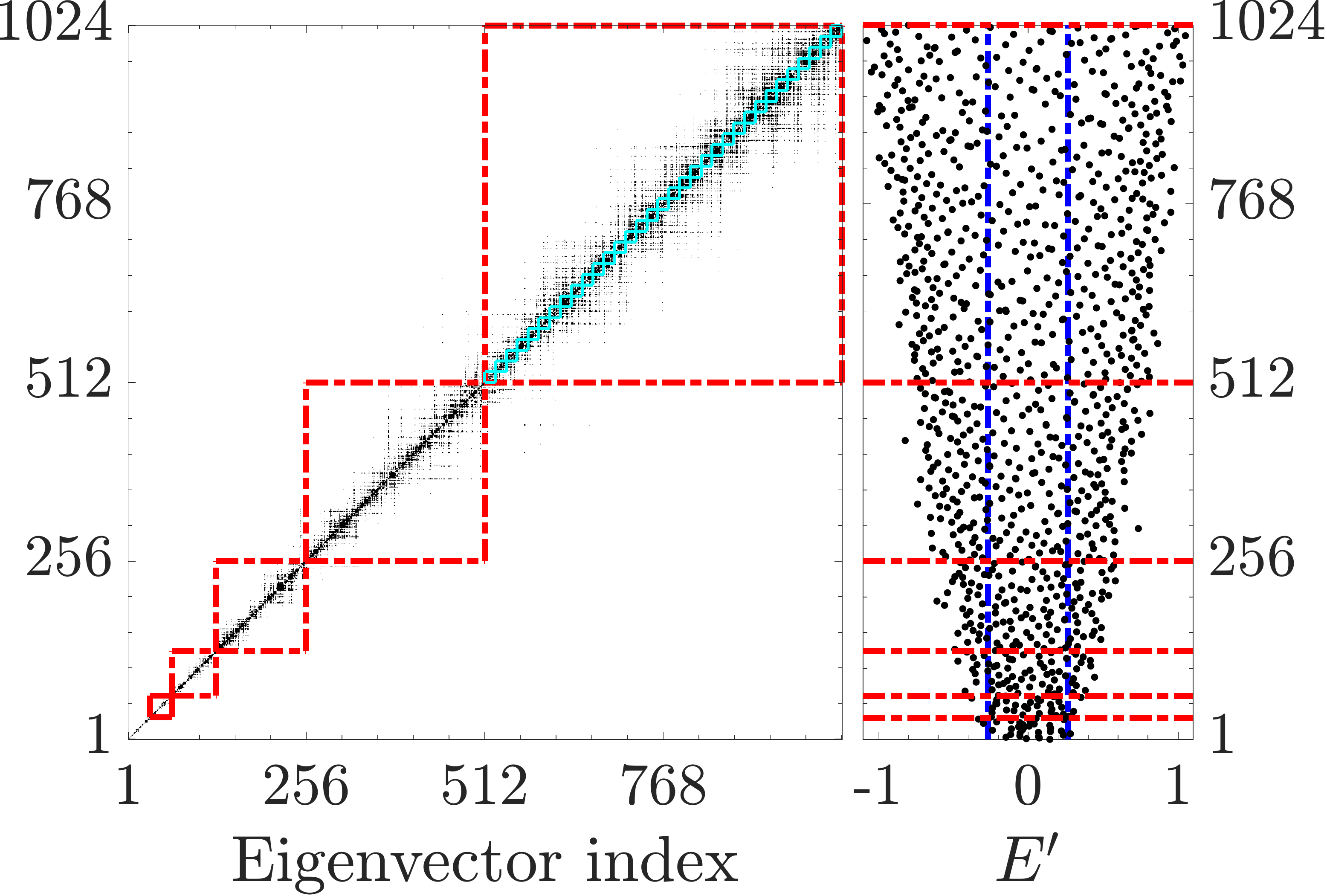}
	\caption{(left)~Threshold-filtered covariance matrix $\tilde{M}$, Eq.~\eqref{eq_covariance}, for $N = 1024$, $\gamma = 0.5$, and $\delta M = 0.5$
		The red dashed lines denote spatial blocks of the form $\Delta_l$ (Eq.~\eqref{eq_part}) with $\delta\zeta = 0.1$ and $l = 0, 1,\dots,5$, while
		the cyan lines show $N^\gamma$ spatial sub-blocks in the last block.
		(right)~Energies of spatially ordered eigenstates. The blue dashed vertical lines denote $\del{-\eg', \eg'}$, i.e.~energy bound of localized states.
	}
	\label{fig_partition}
\end{figure}
Firstly, $\mathcal{O}(N^\gamma)$ states, confined in the first spatial block $\Delta_0$ are all localized with $\xi_0 \sim N^{\delta\zeta}\sim \mathcal{O}(1)$ ($\delta\zeta\simeq \frac{O(1)}{\log N}$). 
Eq.~\eqref{eq_block_DOS} also justifies that such $\delta$-localized states have an energy bandwidth $\eg \approx \mathcal{O}(1)$, not scaling with $N$~\cite{supple}.

Secondly, the eigenstates in the last largest block 
are least localized with a localization length $\xi\sim N^{1-\gamma}$.
This block contains a finite fraction of all sites, $N_l\sim \mathcal{O}(N)$ and then defines the typical fractal dimension $D_2^\text{typ} = 1-\gamma$
in the NEE phase of \bte~\cite{Das2}.
As both the number of eigenstates, with localization length $1\leq \xi_l\leq N^{1-\gamma}$, and the bandwidth, Eq.~\eqref{eq_block_DOS}, increase with $l$,
the distribution of any localization measure exhibits a fat tail~\cite{supple}.
Thus, within $(-\eg, \eg)$, where all the bands overlap, all the localization lengths are possible.
This structure of spatial-separated states with different $\xi_l$ explains the absence of ME and the coexistense of localized and extended states in \bte.


Thirdly, the Anderson equivalence explains the anomalous long-range energy correlations in the NEE phase of \bte.
The eigenvalues from all blocks constitute the global DOS, hence the bandwidth $\epsilon_\beta$ is given by that of the largest block at $l_{\max} \simeq \frac{1-\gamma}{\delta\zeta}$ with a bandwidth $N^{\frac{1-\gamma}{2}}$.
Thus, global mean level spacing is given by $\delta \sim \frac{N^{\frac{1-\gamma}{2}}}{N} = N^{-\frac{1+\gamma}{2}}$.
Contrarily, the smallest level spacing, locally in a sub-block is $\delta_\text{min}=\delta_{l_{\max}}\sim N^{-\frac{1-\gamma}{2}}$.
As $\delta_\text{min}>\delta$, neighboring eigenvalues come from different sub-blocks and are, thus, uncorrelated, while the correlated ones have at least the energy difference $\delta_{\min}$.
The unfolding procedure
rescales $\delta\to 1$, setting a critical dimensionless energy ${\delta_\text{min}}/{\delta} = N^\gamma$, in agreement with numerics.
Any two unfolded energy levels $\cale_{1, 2}$ are uncorrelated if $|\cale_1 - \cale_2|\lesssim N^\gamma$ being from different sub-blocks, while for $|\cale_1 - \cale_2|\gtrsim N^\gamma$
(actual level spacing $\gtrapprox N^{\frac{\gamma - 1}{2}}$), they may belong to the same sub-block and be correlated.
This explains the origin of the anomalous behavior in the power spectrum of the energy fluctuations in the \bte, Eq.~\eqref{eq_P_d_w_whole}.

Finally, we numerically confirm that the eigenstates of \bte\ in the NEE phase are exponentially decaying using the metric defined in \cite{Prosen1, supple}. Therefore we can order the eigenstates according to their localization centers instead of energy. To understand the correlation among such spatially ordered eigenstates, we look at the covariance matrix
\begin{align}
	\label{eq_covariance}
	M_{ij} = \sum_{k = 1}^{N} |\Psi_i(k)\Psi_j(k)| = \begin{cases}
		1, & i = j\\
		\dfrac{2}{\pi}, & \text{ergodic}\\
		\to 0, & \text{localized}
	\end{cases}.
\end{align}
The covariance matrix gives a rather complete idea about the Hilbert space structure.
By plotting the threshold-filtered covariance matrix $\tilde{M}_{ij} = \Theta(M_{ij} - \delta M)$, with the Heaviside step function $\Theta(x)$ and
a threshold $\delta M<\frac{2}{\pi}$, we unveil the eigenstate spatial correlation structure. $\tilde{M}_{ij} = 1$ implies that $i$th and $j$th states have a high degree of overlap, i.e.~they are hybridizing and vice-a-versa. In the ergodic phase $\tilde{M}$ is a dense matrix, while it is sparse in the localized regime.
In \bte, see Fig.~\ref{fig_partition}, $\tilde M$ shows banded structure with the spatial band, increasing with the indices $i,j$, confirming the analytical picture of the block $\Delta_l$.
We have also shown the energy levels of the spatially ordered eigenstates. This further shows in the \bte\ the localized and the NEE states can appear at nearly same energy eventually leading to coexistence in the thermodynamic limit.

To sum up, in this letter, we provide the set of main principles on how to avoid the mobility-edge emergence in short-range disordered models and illustrate them in a well-known example of \bte. With various spectral and localization measures, we uncover the structure of the coexistence of localized and extended states in such model
and confirm these results analytically by the spatially local mapping to the 1D Anderson model with system-size-dependent hopping.
The general principles, formulated in this work and verified on \bte, allow one to realize the coexistence of the localized and extended states
in the same energy interval without fine-tuning which is robust against perturbations and disorder realizations.
Such systems can be used for quantum memory and fault-tolerant quantum calculations, where the localized states, decoupled from the extended modes of the bath, are free from decoherence.
As an outlook, it would be of particular interest to find many-body realizations of the above concepts.

{\it Acknowledgment.\textemdash}A.~K.~D. is supported by an INSPIRE Fellowship, DST, India.
I.~M.~K. acknowledges the support of the Russian Science Foundation, Grant No. 21-12-00409.

\begin{thebibliography}{60}%
	\makeatletter
	\providecommand \@ifxundefined [1]{%
		\@ifx{#1\undefined}
	}%
	\providecommand \@ifnum [1]{%
		\ifnum #1\expandafter \@firstoftwo
		\else \expandafter \@secondoftwo
		\fi
	}%
	\providecommand \@ifx [1]{%
		\ifx #1\expandafter \@firstoftwo
		\else \expandafter \@secondoftwo
		\fi
	}%
	\providecommand \natexlab [1]{#1}%
	\providecommand \enquote  [1]{``#1''}%
	\providecommand \bibnamefont  [1]{#1}%
	\providecommand \bibfnamefont [1]{#1}%
	\providecommand \citenamefont [1]{#1}%
	\providecommand \href@noop [0]{\@secondoftwo}%
	\providecommand \href [0]{\begingroup \@sanitize@url \@href}%
	\providecommand \@href[1]{\@@startlink{#1}\@@href}%
	\providecommand \@@href[1]{\endgroup#1\@@endlink}%
	\providecommand \@sanitize@url [0]{\catcode `\\12\catcode `\$12\catcode
		`\&12\catcode `\#12\catcode `\^12\catcode `\_12\catcode `\%12\relax}%
	\providecommand \@@startlink[1]{}%
	\providecommand \@@endlink[0]{}%
	\providecommand \url  [0]{\begingroup\@sanitize@url \@url }%
	\providecommand \@url [1]{\endgroup\@href {#1}{\urlprefix }}%
	\providecommand \urlprefix  [0]{URL }%
	\providecommand \Eprint [0]{\href }%
	\providecommand \doibase [0]{https://doi.org/}%
	\providecommand \selectlanguage [0]{\@gobble}%
	\providecommand \bibinfo  [0]{\@secondoftwo}%
	\providecommand \bibfield  [0]{\@secondoftwo}%
	\providecommand \translation [1]{[#1]}%
	\providecommand \BibitemOpen [0]{}%
	\providecommand \bibitemStop [0]{}%
	\providecommand \bibitemNoStop [0]{.\EOS\space}%
	\providecommand \EOS [0]{\spacefactor3000\relax}%
	\providecommand \BibitemShut  [1]{\csname bibitem#1\endcsname}%
	\let\auto@bib@innerbib\@empty
	\bibitem [{\citenamefont {Mott}(1967)}]{Mott1967}%
	\BibitemOpen
	\bibfield  {author} {\bibinfo {author} {\bibfnamefont {N.}~\bibnamefont
			{Mott}},\ }\bibfield  {title} {\bibinfo {title} {Electrons in disordered
			structures},\ }\href {https://doi.org/10.1080/00018736700101265} {\bibfield
		{journal} {\bibinfo  {journal} {Advances in Physics}\ }\textbf {\bibinfo
			{volume} {16}},\ \bibinfo {pages} {49} (\bibinfo {year} {1967})}\BibitemShut
	{NoStop}%
	\bibitem [{\citenamefont {Anderson}(1958)}]{Anderson1}%
	\BibitemOpen
	\bibfield  {author} {\bibinfo {author} {\bibfnamefont {P.~W.}\ \bibnamefont
			{Anderson}},\ }\bibfield  {title} {\bibinfo {title} {Absence of diffusion in
			certain random lattices},\ }\href {https://doi.org/10.1103/PhysRev.109.1492}
	{\bibfield  {journal} {\bibinfo  {journal} {Phys. Rev.}\ }\textbf {\bibinfo
			{volume} {109}},\ \bibinfo {pages} {1492} (\bibinfo {year}
		{1958})}\BibitemShut {NoStop}%
	\bibitem [{\citenamefont {Luitz}\ \emph {et~al.}(2015)\citenamefont {Luitz},
		\citenamefont {Laflorencie},\ and\ \citenamefont {Alet}}]{Luitz1}%
	\BibitemOpen
	\bibfield  {author} {\bibinfo {author} {\bibfnamefont {D.~J.}\ \bibnamefont
			{Luitz}}, \bibinfo {author} {\bibfnamefont {N.}~\bibnamefont {Laflorencie}},\
		and\ \bibinfo {author} {\bibfnamefont {F.}~\bibnamefont {Alet}},\ }\bibfield
	{title} {\bibinfo {title} {Many-body localization edge in the random-field
			heisenberg chain},\ }\href {https://doi.org/10.1103/PhysRevB.91.081103}
	{\bibfield  {journal} {\bibinfo  {journal} {Phys. Rev. B}\ }\textbf {\bibinfo
			{volume} {91}},\ \bibinfo {pages} {081103} (\bibinfo {year}
		{2015})}\BibitemShut {NoStop}%
	\bibitem [{\citenamefont {Dumitriu}\ and\ \citenamefont
		{Edelman}(2002)}]{Dumitriu1}%
	\BibitemOpen
	\bibfield  {author} {\bibinfo {author} {\bibfnamefont {I.}~\bibnamefont
			{Dumitriu}}\ and\ \bibinfo {author} {\bibfnamefont {A.}~\bibnamefont
			{Edelman}},\ }\bibfield  {title} {\bibinfo {title} {Matrix models for beta
			ensembles},\ }\href {https://doi.org/10.1063/1.1507823} {\bibfield  {journal}
		{\bibinfo  {journal} {Journal of Mathematical Physics}\ }\textbf {\bibinfo
			{volume} {43}},\ \bibinfo {pages} {5830} (\bibinfo {year} {2002})},\ \Eprint
	{https://arxiv.org/abs/https://doi.org/10.1063/1.1507823}
	{https://doi.org/10.1063/1.1507823} \BibitemShut {NoStop}%
	\bibitem [{\citenamefont {Mehta}(2004)}]{Mehta1}%
	\BibitemOpen
	\bibfield  {author} {\bibinfo {author} {\bibfnamefont {M.}~\bibnamefont
			{Mehta}},\ }\href {https://books.google.co.in/books?id=Fe6uoQEACAAJ} {\emph
		{\bibinfo {title} {Random Matrices}}},\ Pure and Applied Mathematics\
	(\bibinfo  {publisher} {Elsevier Science},\ \bibinfo {year}
	{2004})\BibitemShut {NoStop}%
	\bibitem [{\citenamefont {Berry}\ and\ \citenamefont {Tabor}(1977)}]{Berry3}%
	\BibitemOpen
	\bibfield  {author} {\bibinfo {author} {\bibfnamefont {M.}~\bibnamefont
			{Berry}}\ and\ \bibinfo {author} {\bibfnamefont {M.}~\bibnamefont {Tabor}},\
	}\bibfield  {title} {\bibinfo {title} {Level clustering in the regular
			spectrum},\ }\href {https://doi.org/10.1098/rspa.1977.0140} {\bibfield
		{journal} {\bibinfo  {journal} {Proceedings of the Royal Society of London A:
				Mathematical, Physical and Engineering Sciences}\ }\textbf {\bibinfo {volume}
			{356}},\ \bibinfo {pages} {375} (\bibinfo {year} {1977})}\BibitemShut
	{NoStop}%
	\bibitem [{\citenamefont {Bohigas}\ \emph {et~al.}(1984)\citenamefont
		{Bohigas}, \citenamefont {Giannoni},\ and\ \citenamefont
		{Schmit}}]{Bohigas1}%
	\BibitemOpen
	\bibfield  {author} {\bibinfo {author} {\bibfnamefont {O.}~\bibnamefont
			{Bohigas}}, \bibinfo {author} {\bibfnamefont {M.~J.}\ \bibnamefont
			{Giannoni}},\ and\ \bibinfo {author} {\bibfnamefont {C.}~\bibnamefont
			{Schmit}},\ }\bibfield  {title} {\bibinfo {title} {Characterization of
			chaotic quantum spectra and universality of level fluctuation laws},\ }\href
	{https://doi.org/10.1103/PhysRevLett.52.1} {\bibfield  {journal} {\bibinfo
			{journal} {Phys. Rev. Lett.}\ }\textbf {\bibinfo {volume} {52}},\ \bibinfo
		{pages} {1} (\bibinfo {year} {1984})}\BibitemShut {NoStop}%
	\bibitem [{\citenamefont {De~Luca}\ \emph {et~al.}(2014)\citenamefont
		{De~Luca}, \citenamefont {Altshuler}, \citenamefont {Kravtsov},\ and\
		\citenamefont {Scardicchio}}]{Luca1}%
	\BibitemOpen
	\bibfield  {author} {\bibinfo {author} {\bibfnamefont {A.}~\bibnamefont
			{De~Luca}}, \bibinfo {author} {\bibfnamefont {B.~L.}\ \bibnamefont
			{Altshuler}}, \bibinfo {author} {\bibfnamefont {V.~E.}\ \bibnamefont
			{Kravtsov}},\ and\ \bibinfo {author} {\bibfnamefont {A.}~\bibnamefont
			{Scardicchio}},\ }\bibfield  {title} {\bibinfo {title} {Anderson localization
			on the bethe lattice: Nonergodicity of extended states},\ }\href
	{https://doi.org/10.1103/PhysRevLett.113.046806} {\bibfield  {journal}
		{\bibinfo  {journal} {Phys. Rev. Lett.}\ }\textbf {\bibinfo {volume} {113}},\
		\bibinfo {pages} {046806} (\bibinfo {year} {2014})}\BibitemShut {NoStop}%
	\bibitem [{\citenamefont {Garc\'ia-Mata}\ \emph {et~al.}(2017)\citenamefont
		{Garc\'ia-Mata}, \citenamefont {Giraud}, \citenamefont {Georgeot},
		\citenamefont {Martin}, \citenamefont {Dubertrand},\ and\ \citenamefont
		{Lemari\'e}}]{Garcia2}%
	\BibitemOpen
	\bibfield  {author} {\bibinfo {author} {\bibfnamefont {I.}~\bibnamefont
			{Garc\'ia-Mata}}, \bibinfo {author} {\bibfnamefont {O.}~\bibnamefont
			{Giraud}}, \bibinfo {author} {\bibfnamefont {B.}~\bibnamefont {Georgeot}},
		\bibinfo {author} {\bibfnamefont {J.}~\bibnamefont {Martin}}, \bibinfo
		{author} {\bibfnamefont {R.}~\bibnamefont {Dubertrand}},\ and\ \bibinfo
		{author} {\bibfnamefont {G.}~\bibnamefont {Lemari\'e}},\ }\bibfield  {title}
	{\bibinfo {title} {Scaling theory of the anderson transition in random
			graphs: Ergodicity and universality},\ }\href
	{https://doi.org/10.1103/PhysRevLett.118.166801} {\bibfield  {journal}
		{\bibinfo  {journal} {Phys. Rev. Lett.}\ }\textbf {\bibinfo {volume} {118}},\
		\bibinfo {pages} {166801} (\bibinfo {year} {2017})}\BibitemShut {NoStop}%
	\bibitem [{\citenamefont {Ray}\ \emph {et~al.}(2018)\citenamefont {Ray},
		\citenamefont {Ghosh},\ and\ \citenamefont {Sinha}}]{Ray1}%
	\BibitemOpen
	\bibfield  {author} {\bibinfo {author} {\bibfnamefont {S.}~\bibnamefont
			{Ray}}, \bibinfo {author} {\bibfnamefont {A.}~\bibnamefont {Ghosh}},\ and\
		\bibinfo {author} {\bibfnamefont {S.}~\bibnamefont {Sinha}},\ }\bibfield
	{title} {\bibinfo {title} {Drive-induced delocalization in the aubry-andr\'e
			model},\ }\href {https://doi.org/10.1103/PhysRevE.97.010101} {\bibfield
		{journal} {\bibinfo  {journal} {Phys. Rev. E}\ }\textbf {\bibinfo {volume}
			{97}},\ \bibinfo {pages} {010101} (\bibinfo {year} {2018})}\BibitemShut
	{NoStop}%
	\bibitem [{\citenamefont {Wang}\ \emph {et~al.}(2022)\citenamefont {Wang},
		\citenamefont {Zhang}, \citenamefont {Sun}, \citenamefont {Poon},\ and\
		\citenamefont {Liu}}]{Wang3}%
	\BibitemOpen
	\bibfield  {author} {\bibinfo {author} {\bibfnamefont {Y.}~\bibnamefont
			{Wang}}, \bibinfo {author} {\bibfnamefont {L.}~\bibnamefont {Zhang}},
		\bibinfo {author} {\bibfnamefont {W.}~\bibnamefont {Sun}}, \bibinfo {author}
		{\bibfnamefont {T.-F.~J.}\ \bibnamefont {Poon}},\ and\ \bibinfo {author}
		{\bibfnamefont {X.-J.}\ \bibnamefont {Liu}},\ }\bibfield  {title} {\bibinfo
		{title} {Quantum phase with coexisting localized, extended, and critical
			zones},\ }\href {https://doi.org/10.1103/PhysRevB.106.L140203} {\bibfield
		{journal} {\bibinfo  {journal} {Phys. Rev. B}\ }\textbf {\bibinfo {volume}
			{106}},\ \bibinfo {pages} {L140203} (\bibinfo {year} {2022})}\BibitemShut
	{NoStop}%
	\bibitem [{\citenamefont {Pino}\ \emph {et~al.}(2017)\citenamefont {Pino},
		\citenamefont {Kravtsov}, \citenamefont {Altshuler},\ and\ \citenamefont
		{Ioffe}}]{Pino2}%
	\BibitemOpen
	\bibfield  {author} {\bibinfo {author} {\bibfnamefont {M.}~\bibnamefont
			{Pino}}, \bibinfo {author} {\bibfnamefont {V.~E.}\ \bibnamefont {Kravtsov}},
		\bibinfo {author} {\bibfnamefont {B.~L.}\ \bibnamefont {Altshuler}},\ and\
		\bibinfo {author} {\bibfnamefont {L.~B.}\ \bibnamefont {Ioffe}},\ }\bibfield
	{title} {\bibinfo {title} {Multifractal metal in a disordered josephson
			junctions array},\ }\href {https://doi.org/10.1103/PhysRevB.96.214205}
	{\bibfield  {journal} {\bibinfo  {journal} {Phys. Rev. B}\ }\textbf {\bibinfo
			{volume} {96}},\ \bibinfo {pages} {214205} (\bibinfo {year}
		{2017})}\BibitemShut {NoStop}%
	\bibitem [{\citenamefont {Pino}\ \emph {et~al.}(2016)\citenamefont {Pino},
		\citenamefont {Ioffe},\ and\ \citenamefont {Altshuler}}]{Pino3}%
	\BibitemOpen
	\bibfield  {author} {\bibinfo {author} {\bibfnamefont {M.}~\bibnamefont
			{Pino}}, \bibinfo {author} {\bibfnamefont {L.~B.}\ \bibnamefont {Ioffe}},\
		and\ \bibinfo {author} {\bibfnamefont {B.~L.}\ \bibnamefont {Altshuler}},\
	}\bibfield  {title} {\bibinfo {title} {Nonergodic metallic and insulating
			phases of josephson junction chains},\ }\href@noop {} {\bibfield  {journal}
		{\bibinfo  {journal} {Proceedings of the National Academy of Sciences}\
		}\textbf {\bibinfo {volume} {113}},\ \bibinfo {pages} {536} (\bibinfo {year}
		{2016})}\BibitemShut {NoStop}%
	\bibitem [{\citenamefont {Das}\ and\ \citenamefont
		{Ghosh}(2022{\natexlab{a}})}]{Das4}%
	\BibitemOpen
	\bibfield  {author} {\bibinfo {author} {\bibfnamefont {A.~K.}\ \bibnamefont
			{Das}}\ and\ \bibinfo {author} {\bibfnamefont {A.}~\bibnamefont {Ghosh}},\
	}\bibfield  {title} {\bibinfo {title} {Transport in deformed centrosymmetric
			networks},\ }\href {https://doi.org/10.1103/PhysRevE.106.064112} {\bibfield
		{journal} {\bibinfo  {journal} {Phys. Rev. E}\ }\textbf {\bibinfo {volume}
			{106}},\ \bibinfo {pages} {064112} (\bibinfo {year}
		{2022}{\natexlab{a}})}\BibitemShut {NoStop}%
	\bibitem [{\citenamefont {Fyodorov}\ and\ \citenamefont
		{Giraud}(2015)}]{Fyodorov1}%
	\BibitemOpen
	\bibfield  {author} {\bibinfo {author} {\bibfnamefont {Y.~V.}\ \bibnamefont
			{Fyodorov}}\ and\ \bibinfo {author} {\bibfnamefont {O.}~\bibnamefont
			{Giraud}},\ }\bibfield  {title} {\bibinfo {title} {High values of
			disorder-generated multifractals and logarithmically correlated processes},\
	}\href {https://doi.org/https://doi.org/10.1016/j.chaos.2014.11.018}
	{\bibfield  {journal} {\bibinfo  {journal} {Chaos, Solitons \& Fractals}\
		}\textbf {\bibinfo {volume} {74}},\ \bibinfo {pages} {15} (\bibinfo {year}
		{2015})},\ \bibinfo {note} {extreme Events and its Applications}\BibitemShut
	{NoStop}%
	\bibitem [{\citenamefont {Kravtsov}\ \emph {et~al.}(2015)\citenamefont
		{Kravtsov}, \citenamefont {Khaymovich}, \citenamefont {Cuevas},\ and\
		\citenamefont {Amini}}]{Kravtsov1}%
	\BibitemOpen
	\bibfield  {author} {\bibinfo {author} {\bibfnamefont {V.~E.}\ \bibnamefont
			{Kravtsov}}, \bibinfo {author} {\bibfnamefont {I.~M.}\ \bibnamefont
			{Khaymovich}}, \bibinfo {author} {\bibfnamefont {E.}~\bibnamefont {Cuevas}},\
		and\ \bibinfo {author} {\bibfnamefont {M.}~\bibnamefont {Amini}},\ }\bibfield
	{title} {\bibinfo {title} {A random matrix model with localization and
			ergodic transitions},\ }\href
	{https://doi.org/10.1088/1367-2630/17/12/122002} {\bibfield  {journal}
		{\bibinfo  {journal} {New Journal of Physics}\ }\textbf {\bibinfo {volume}
			{17}},\ \bibinfo {pages} {122002} (\bibinfo {year} {2015})}\BibitemShut
	{NoStop}%
	\bibitem [{\citenamefont {Nosov}\ \emph {et~al.}(2019)\citenamefont {Nosov},
		\citenamefont {Khaymovich},\ and\ \citenamefont
		{Kravtsov}}]{Nosov2019correlation}%
	\BibitemOpen
	\bibfield  {author} {\bibinfo {author} {\bibfnamefont {P.~A.}\ \bibnamefont
			{Nosov}}, \bibinfo {author} {\bibfnamefont {I.~M.}\ \bibnamefont
			{Khaymovich}},\ and\ \bibinfo {author} {\bibfnamefont {V.~E.}\ \bibnamefont
			{Kravtsov}},\ }\bibfield  {title} {\bibinfo {title} {Correlation-induced
			localization},\ }\href {https://doi.org/10.1103/PhysRevB.99.104203}
	{\bibfield  {journal} {\bibinfo  {journal} {Physical Review B}\ }\textbf
		{\bibinfo {volume} {99}},\ \bibinfo {pages} {104203} (\bibinfo {year}
		{2019})}\BibitemShut {NoStop}%
	\bibitem [{\citenamefont {Nosov}\ and\ \citenamefont
		{Khaymovich}(2019)}]{Nosov2019mixtures}%
	\BibitemOpen
	\bibfield  {author} {\bibinfo {author} {\bibfnamefont {P.~A.}\ \bibnamefont
			{Nosov}}\ and\ \bibinfo {author} {\bibfnamefont {I.~M.}\ \bibnamefont
			{Khaymovich}},\ }\bibfield  {title} {\bibinfo {title} {Robustness of
			delocalization to the inclusion of soft constraints in long-range random
			models},\ }\href {https://doi.org/10.1103/PhysRevB.99.224208} {\bibfield
		{journal} {\bibinfo  {journal} {Phys. Rev. B}\ }\textbf {\bibinfo {volume}
			{99}},\ \bibinfo {pages} {224208} (\bibinfo {year} {2019})}\BibitemShut
	{NoStop}%
	\bibitem [{\citenamefont {Khaymovich}\ \emph {et~al.}(2020)\citenamefont
		{Khaymovich}, \citenamefont {Kravtsov}, \citenamefont {Altshuler},\ and\
		\citenamefont {Ioffe}}]{Khaymovich1}%
	\BibitemOpen
	\bibfield  {author} {\bibinfo {author} {\bibfnamefont {I.~M.}\ \bibnamefont
			{Khaymovich}}, \bibinfo {author} {\bibfnamefont {V.~E.}\ \bibnamefont
			{Kravtsov}}, \bibinfo {author} {\bibfnamefont {B.~L.}\ \bibnamefont
			{Altshuler}},\ and\ \bibinfo {author} {\bibfnamefont {L.~B.}\ \bibnamefont
			{Ioffe}},\ }\bibfield  {title} {\bibinfo {title} {Fragile extended phases in
			the log-normal rosenzweig-porter model},\ }\href
	{https://doi.org/10.1103/PhysRevResearch.2.043346} {\bibfield  {journal}
		{\bibinfo  {journal} {Phys. Rev. Research}\ }\textbf {\bibinfo {volume}
			{2}},\ \bibinfo {pages} {043346} (\bibinfo {year} {2020})}\BibitemShut
	{NoStop}%
	\bibitem [{\citenamefont {Cizeau}\ and\ \citenamefont
		{Bouchaud}(1994)}]{Cizeau1}%
	\BibitemOpen
	\bibfield  {author} {\bibinfo {author} {\bibfnamefont {P.}~\bibnamefont
			{Cizeau}}\ and\ \bibinfo {author} {\bibfnamefont {J.~P.}\ \bibnamefont
			{Bouchaud}},\ }\bibfield  {title} {\bibinfo {title} {Theory of l\'evy
			matrices},\ }\href {https://doi.org/10.1103/PhysRevE.50.1810} {\bibfield
		{journal} {\bibinfo  {journal} {Phys. Rev. E}\ }\textbf {\bibinfo {volume}
			{50}},\ \bibinfo {pages} {1810} (\bibinfo {year} {1994})}\BibitemShut
	{NoStop}%
	\bibitem [{\citenamefont {Das}\ and\ \citenamefont
		{Ghosh}(2022{\natexlab{b}})}]{Das3}%
	\BibitemOpen
	\bibfield  {author} {\bibinfo {author} {\bibfnamefont {A.~K.}\ \bibnamefont
			{Das}}\ and\ \bibinfo {author} {\bibfnamefont {A.}~\bibnamefont {Ghosh}},\
	}\bibfield  {title} {\bibinfo {title} {Chaos due to symmetry-breaking in
			deformed poisson ensemble},\ }\href@noop {} {\bibfield  {journal} {\bibinfo
			{journal} {Journal of Statistical Mechanics: Theory and Experiment}\ }\textbf
		{\bibinfo {volume} {2022}},\ \bibinfo {pages} {063101} (\bibinfo {year}
		{2022}{\natexlab{b}})}\BibitemShut {NoStop}%
	\bibitem [{\citenamefont {Kravtsov}\ \emph {et~al.}(2020)\citenamefont
		{Kravtsov}, \citenamefont {Khaymovich}, \citenamefont {Altshuler},\ and\
		\citenamefont {Ioffe}}]{LN-RP_RRG}%
	\BibitemOpen
	\bibfield  {author} {\bibinfo {author} {\bibfnamefont {V.~E.}\ \bibnamefont
			{Kravtsov}}, \bibinfo {author} {\bibfnamefont {I.~M.}\ \bibnamefont
			{Khaymovich}}, \bibinfo {author} {\bibfnamefont {B.~L.}\ \bibnamefont
			{Altshuler}},\ and\ \bibinfo {author} {\bibfnamefont {L.~B.}\ \bibnamefont
			{Ioffe}},\ }\bibfield  {title} {\bibinfo {title} {Localization transition on
			the random regular graph as an unstable tricritical point in a log-normal
			{Rosenzweig}-{Porter} random matrix ensemble},\ }\Eprint
	{https://arxiv.org/abs/2002.02979} {arXiv:2002.02979}  (\bibinfo {year}
	{2020})\BibitemShut {NoStop}%
	\bibitem [{\citenamefont {Khaymovich}\ and\ \citenamefont
		{Kravtsov}(2021)}]{LN-RP_dyn}%
	\BibitemOpen
	\bibfield  {author} {\bibinfo {author} {\bibfnamefont {I.~M.}\ \bibnamefont
			{Khaymovich}}\ and\ \bibinfo {author} {\bibfnamefont {V.~E.}\ \bibnamefont
			{Kravtsov}},\ }\bibfield  {title} {\bibinfo {title} {Dynamical phases in a
			``multifractal'' {Rosenzweig}-{Porter} model},\ }\href
	{https://doi.org/10.21468/SciPostPhys.11.2.045} {\bibfield  {journal}
		{\bibinfo  {journal} {SciPost Phys.}\ }\textbf {\bibinfo {volume} {11}},\
		\bibinfo {pages} {45} (\bibinfo {year} {2021})}\BibitemShut {NoStop}%
	\bibitem [{\citenamefont {Kutlin}\ and\ \citenamefont
		{Khaymovich}(2021)}]{Kutlin2021emergent}%
	\BibitemOpen
	\bibfield  {author} {\bibinfo {author} {\bibfnamefont {A.~G.}\ \bibnamefont
			{Kutlin}}\ and\ \bibinfo {author} {\bibfnamefont {I.~M.}\ \bibnamefont
			{Khaymovich}},\ }\bibfield  {title} {\bibinfo {title} {Emergent fractal phase
			in energy stratified random models},\ }\href
	{https://doi.org/10.21468/SciPostPhys.11.6.101} {\bibfield  {journal}
		{\bibinfo  {journal} {SciPost Phys.}\ }\textbf {\bibinfo {volume} {11}},\
		\bibinfo {pages} {101} (\bibinfo {year} {2021})}\BibitemShut {NoStop}%
	\bibitem [{\citenamefont {Tang}\ and\ \citenamefont
		{Khaymovich}(2022)}]{Tang2022nonergodic}%
	\BibitemOpen
	\bibfield  {author} {\bibinfo {author} {\bibfnamefont {W.}~\bibnamefont
			{Tang}}\ and\ \bibinfo {author} {\bibfnamefont {I.~M.}\ \bibnamefont
			{Khaymovich}},\ }\bibfield  {title} {\bibinfo {title} {Non-ergodic
			delocalized phase with {P}oisson level statistics},\ }\href
	{https://doi.org/10.22331/q-2022-06-09-733} {\bibfield  {journal} {\bibinfo
			{journal} {{Quantum}}\ }\textbf {\bibinfo {volume} {6}},\ \bibinfo {pages}
		{733} (\bibinfo {year} {2022})}\BibitemShut {NoStop}%
	\bibitem [{\citenamefont {Motamarri}\ \emph {et~al.}(2021)\citenamefont
		{Motamarri}, \citenamefont {Gorsky},\ and\ \citenamefont
		{Khaymovich}}]{Motamarri2021RDM}%
	\BibitemOpen
	\bibfield  {author} {\bibinfo {author} {\bibfnamefont {V.}~\bibnamefont
			{Motamarri}}, \bibinfo {author} {\bibfnamefont {A.~S.}\ \bibnamefont
			{Gorsky}},\ and\ \bibinfo {author} {\bibfnamefont {I.~M.}\ \bibnamefont
			{Khaymovich}},\ }\bibfield  {title} {\bibinfo {title} {Localization and
			fractality in disordered russian doll model},\ }\Eprint
	{https://arxiv.org/abs/2112.05066} {arXiv:2112.05066}  (\bibinfo {year}
	{2021})\BibitemShut {NoStop}%
	\bibitem [{\citenamefont {Tikhonov}\ and\ \citenamefont
		{Mirlin}(2016)}]{Tikhonov2}%
	\BibitemOpen
	\bibfield  {author} {\bibinfo {author} {\bibfnamefont {K.~S.}\ \bibnamefont
			{Tikhonov}}\ and\ \bibinfo {author} {\bibfnamefont {A.~D.}\ \bibnamefont
			{Mirlin}},\ }\bibfield  {title} {\bibinfo {title} {Fractality of wave
			functions on a cayley tree: Difference between tree and locally treelike
			graph without boundary},\ }\href {https://doi.org/10.1103/PhysRevB.94.184203}
	{\bibfield  {journal} {\bibinfo  {journal} {Phys. Rev. B}\ }\textbf {\bibinfo
			{volume} {94}},\ \bibinfo {pages} {184203} (\bibinfo {year}
		{2016})}\BibitemShut {NoStop}%
	\bibitem [{\citenamefont {Luitz}\ \emph {et~al.}(2020)\citenamefont {Luitz},
		\citenamefont {Khaymovich},\ and\ \citenamefont {Lev}}]{Luitz3}%
	\BibitemOpen
	\bibfield  {author} {\bibinfo {author} {\bibfnamefont {D.~J.}\ \bibnamefont
			{Luitz}}, \bibinfo {author} {\bibfnamefont {I.~M.}\ \bibnamefont
			{Khaymovich}},\ and\ \bibinfo {author} {\bibfnamefont {Y.~B.}\ \bibnamefont
			{Lev}},\ }\bibfield  {title} {\bibinfo {title} {{Multifractality and its role
				in anomalous transport in the disordered XXZ spin-chain}},\ }\href
	{https://doi.org/10.21468/SciPostPhysCore.2.2.006} {\bibfield  {journal}
		{\bibinfo  {journal} {SciPost Phys. Core}\ }\textbf {\bibinfo {volume} {2}},\
		\bibinfo {pages} {006} (\bibinfo {year} {2020})}\BibitemShut {NoStop}%
	\bibitem [{\citenamefont {Das}\ and\ \citenamefont
		{Ghosh}(2022{\natexlab{c}})}]{Das2}%
	\BibitemOpen
	\bibfield  {author} {\bibinfo {author} {\bibfnamefont {A.~K.}\ \bibnamefont
			{Das}}\ and\ \bibinfo {author} {\bibfnamefont {A.}~\bibnamefont {Ghosh}},\
	}\bibfield  {title} {\bibinfo {title} {Nonergodic extended states in the
			$\ensuremath{\beta}$ ensemble},\ }\href
	{https://doi.org/10.1103/PhysRevE.105.054121} {\bibfield  {journal} {\bibinfo
			{journal} {Phys. Rev. E}\ }\textbf {\bibinfo {volume} {105}},\ \bibinfo
		{pages} {054121} (\bibinfo {year} {2022}{\natexlab{c}})}\BibitemShut
	{NoStop}%
	\bibitem [{\citenamefont {Rosenzweig}\ and\ \citenamefont
		{Porter}(1960)}]{Rosenzweig1}%
	\BibitemOpen
	\bibfield  {author} {\bibinfo {author} {\bibfnamefont {N.}~\bibnamefont
			{Rosenzweig}}\ and\ \bibinfo {author} {\bibfnamefont {C.~E.}\ \bibnamefont
			{Porter}},\ }\bibfield  {title} {\bibinfo {title} {Repulsion of energy levels
			in complex atomic spectra},\ }\href
	{https://doi.org/10.1103/PhysRev.120.1698} {\bibfield  {journal} {\bibinfo
			{journal} {Phys. Rev.}\ }\textbf {\bibinfo {volume} {120}},\ \bibinfo {pages}
		{1698} (\bibinfo {year} {1960})}\BibitemShut {NoStop}%
	\bibitem [{sup()}]{supple}%
	\BibitemOpen
	\bibinfo {note} {See Supplementary Material}\BibitemShut {NoStop}%
	\bibitem [{\citenamefont {Cuesta}\ and\ \citenamefont
		{S{\'a}nchez}(2004)}]{Cuesta1}%
	\BibitemOpen
	\bibfield  {author} {\bibinfo {author} {\bibfnamefont {J.~A.}\ \bibnamefont
			{Cuesta}}\ and\ \bibinfo {author} {\bibfnamefont {A.}~\bibnamefont
			{S{\'a}nchez}},\ }\bibfield  {title} {\bibinfo {title} {General non-existence
			theorem for phase transitions in one-dimensional systems with short range
			interactions, and physical examples of such transitions},\ }\href
	{https://doi.org/10.1023/b:joss.0000022373.63640.4e} {\bibfield  {journal}
		{\bibinfo  {journal} {Journal of statistical physics}\ }\textbf {\bibinfo
			{volume} {115}},\ \bibinfo {pages} {869} (\bibinfo {year}
		{2004})}\BibitemShut {NoStop}%
	\bibitem [{\citenamefont {Albert}\ and\ \citenamefont
		{Leboeuf}(2010)}]{Albert1}%
	\BibitemOpen
	\bibfield  {author} {\bibinfo {author} {\bibfnamefont {M.}~\bibnamefont
			{Albert}}\ and\ \bibinfo {author} {\bibfnamefont {P.}~\bibnamefont
			{Leboeuf}},\ }\bibfield  {title} {\bibinfo {title} {Localization by
			bichromatic potentials versus anderson localization},\ }\href
	{https://doi.org/10.1103/PhysRevA.81.013614} {\bibfield  {journal} {\bibinfo
			{journal} {Phys. Rev. A}\ }\textbf {\bibinfo {volume} {81}},\ \bibinfo
		{pages} {013614} (\bibinfo {year} {2010})}\BibitemShut {NoStop}%
	\bibitem [{\citenamefont {Metz}\ \emph {et~al.}(2010)\citenamefont {Metz},
		\citenamefont {Neri},\ and\ \citenamefont {Boll\'e}}]{Metz3}%
	\BibitemOpen
	\bibfield  {author} {\bibinfo {author} {\bibfnamefont {F.~L.}\ \bibnamefont
			{Metz}}, \bibinfo {author} {\bibfnamefont {I.}~\bibnamefont {Neri}},\ and\
		\bibinfo {author} {\bibfnamefont {D.}~\bibnamefont {Boll\'e}},\ }\bibfield
	{title} {\bibinfo {title} {Localization transition in symmetric random
			matrices},\ }\href {https://doi.org/10.1103/PhysRevE.82.031135} {\bibfield
		{journal} {\bibinfo  {journal} {Phys. Rev. E}\ }\textbf {\bibinfo {volume}
			{82}},\ \bibinfo {pages} {031135} (\bibinfo {year} {2010})}\BibitemShut
	{NoStop}%
	\bibitem [{\citenamefont {Gopalakrishnan}(2017)}]{Gopalakrishnan2}%
	\BibitemOpen
	\bibfield  {author} {\bibinfo {author} {\bibfnamefont {S.}~\bibnamefont
			{Gopalakrishnan}},\ }\bibfield  {title} {\bibinfo {title} {Self-dual
			quasiperiodic systems with power-law hopping},\ }\href
	{https://doi.org/10.1103/PhysRevB.96.054202} {\bibfield  {journal} {\bibinfo
			{journal} {Phys. Rev. B}\ }\textbf {\bibinfo {volume} {96}},\ \bibinfo
		{pages} {054202} (\bibinfo {year} {2017})}\BibitemShut {NoStop}%
	\bibitem [{\citenamefont {Deng}\ \emph {et~al.}(2019)\citenamefont {Deng},
		\citenamefont {Ray}, \citenamefont {Sinha}, \citenamefont {Shlyapnikov},\
		and\ \citenamefont {Santos}}]{Deng2}%
	\BibitemOpen
	\bibfield  {author} {\bibinfo {author} {\bibfnamefont {X.}~\bibnamefont
			{Deng}}, \bibinfo {author} {\bibfnamefont {S.}~\bibnamefont {Ray}}, \bibinfo
		{author} {\bibfnamefont {S.}~\bibnamefont {Sinha}}, \bibinfo {author}
		{\bibfnamefont {G.~V.}\ \bibnamefont {Shlyapnikov}},\ and\ \bibinfo {author}
		{\bibfnamefont {L.}~\bibnamefont {Santos}},\ }\bibfield  {title} {\bibinfo
		{title} {One-dimensional quasicrystals with power-law hopping},\ }\href
	{https://doi.org/10.1103/PhysRevLett.123.025301} {\bibfield  {journal}
		{\bibinfo  {journal} {Phys. Rev. Lett.}\ }\textbf {\bibinfo {volume} {123}},\
		\bibinfo {pages} {025301} (\bibinfo {year} {2019})}\BibitemShut {NoStop}%
	\bibitem [{\citenamefont {Wang}\ \emph {et~al.}(2020)\citenamefont {Wang},
		\citenamefont {Xia}, \citenamefont {Zhang}, \citenamefont {Yao},
		\citenamefont {Chen}, \citenamefont {You}, \citenamefont {Zhou},\ and\
		\citenamefont {Liu}}]{Wang_mosaic}%
	\BibitemOpen
	\bibfield  {author} {\bibinfo {author} {\bibfnamefont {Y.}~\bibnamefont
			{Wang}}, \bibinfo {author} {\bibfnamefont {X.}~\bibnamefont {Xia}}, \bibinfo
		{author} {\bibfnamefont {L.}~\bibnamefont {Zhang}}, \bibinfo {author}
		{\bibfnamefont {H.}~\bibnamefont {Yao}}, \bibinfo {author} {\bibfnamefont
			{S.}~\bibnamefont {Chen}}, \bibinfo {author} {\bibfnamefont {J.}~\bibnamefont
			{You}}, \bibinfo {author} {\bibfnamefont {Q.}~\bibnamefont {Zhou}},\ and\
		\bibinfo {author} {\bibfnamefont {X.-J.}\ \bibnamefont {Liu}},\ }\bibfield
	{title} {\bibinfo {title} {One-dimensional quasiperiodic mosaic lattice with
			exact mobility edges},\ }\href
	{https://doi.org/10.1103/PhysRevLett.125.196604} {\bibfield  {journal}
		{\bibinfo  {journal} {Phys. Rev. Lett.}\ }\textbf {\bibinfo {volume} {125}},\
		\bibinfo {pages} {196604} (\bibinfo {year} {2020})}\BibitemShut {NoStop}%
	\bibitem [{\citenamefont {Das~Sarma}\ \emph {et~al.}(1988)\citenamefont
		{Das~Sarma}, \citenamefont {He},\ and\ \citenamefont {Xie}}]{DasSarma1}%
	\BibitemOpen
	\bibfield  {author} {\bibinfo {author} {\bibfnamefont {S.}~\bibnamefont
			{Das~Sarma}}, \bibinfo {author} {\bibfnamefont {S.}~\bibnamefont {He}},\ and\
		\bibinfo {author} {\bibfnamefont {X.~C.}\ \bibnamefont {Xie}},\ }\bibfield
	{title} {\bibinfo {title} {Mobility edge in a model one-dimensional
			potential},\ }\href {https://doi.org/10.1103/PhysRevLett.61.2144} {\bibfield
		{journal} {\bibinfo  {journal} {Phys. Rev. Lett.}\ }\textbf {\bibinfo
			{volume} {61}},\ \bibinfo {pages} {2144} (\bibinfo {year}
		{1988})}\BibitemShut {NoStop}%
	\bibitem [{\citenamefont {Biddle}\ and\ \citenamefont
		{Das~Sarma}(2010)}]{DasSarma2}%
	\BibitemOpen
	\bibfield  {author} {\bibinfo {author} {\bibfnamefont {J.}~\bibnamefont
			{Biddle}}\ and\ \bibinfo {author} {\bibfnamefont {S.}~\bibnamefont
			{Das~Sarma}},\ }\bibfield  {title} {\bibinfo {title} {Predicted mobility
			edges in one-dimensional incommensurate optical lattices: An exactly solvable
			model of anderson localization},\ }\href
	{https://doi.org/10.1103/PhysRevLett.104.070601} {\bibfield  {journal}
		{\bibinfo  {journal} {Phys. Rev. Lett.}\ }\textbf {\bibinfo {volume} {104}},\
		\bibinfo {pages} {070601} (\bibinfo {year} {2010})}\BibitemShut {NoStop}%
	\bibitem [{\citenamefont {Ganeshan}\ \emph {et~al.}(2015)\citenamefont
		{Ganeshan}, \citenamefont {Pixley},\ and\ \citenamefont
		{Das~Sarma}}]{DasSarma3}%
	\BibitemOpen
	\bibfield  {author} {\bibinfo {author} {\bibfnamefont {S.}~\bibnamefont
			{Ganeshan}}, \bibinfo {author} {\bibfnamefont {J.~H.}\ \bibnamefont
			{Pixley}},\ and\ \bibinfo {author} {\bibfnamefont {S.}~\bibnamefont
			{Das~Sarma}},\ }\bibfield  {title} {\bibinfo {title} {Nearest neighbor tight
			binding models with an exact mobility edge in one dimension},\ }\href
	{https://doi.org/10.1103/PhysRevLett.114.146601} {\bibfield  {journal}
		{\bibinfo  {journal} {Phys. Rev. Lett.}\ }\textbf {\bibinfo {volume} {114}},\
		\bibinfo {pages} {146601} (\bibinfo {year} {2015})}\BibitemShut {NoStop}%
	\bibitem [{\citenamefont {Ahmed}\ \emph {et~al.}(2022)\citenamefont {Ahmed},
		\citenamefont {Ramachandran}, \citenamefont {Khaymovich},\ and\ \citenamefont
		{Sharma}}]{Ahmed2022flat}%
	\BibitemOpen
	\bibfield  {author} {\bibinfo {author} {\bibfnamefont {A.}~\bibnamefont
			{Ahmed}}, \bibinfo {author} {\bibfnamefont {A.}~\bibnamefont {Ramachandran}},
		\bibinfo {author} {\bibfnamefont {I.~M.}\ \bibnamefont {Khaymovich}},\ and\
		\bibinfo {author} {\bibfnamefont {A.}~\bibnamefont {Sharma}},\ }\bibfield
	{title} {\bibinfo {title} {Flat band based multifractality in the
			all-band-flat diamond chain},\ }\href
	{https://doi.org/10.1103/PhysRevB.106.205119} {\bibfield  {journal} {\bibinfo
			{journal} {Phys. Rev. B}\ }\textbf {\bibinfo {volume} {106}},\ \bibinfo
		{pages} {205119} (\bibinfo {year} {2022})}\BibitemShut {NoStop}%
	\bibitem [{\citenamefont {Bulka}\ \emph {et~al.}(1985)\citenamefont {Bulka},
		\citenamefont {Kramer},\ and\ \citenamefont {MacKinnon}}]{Bulka1}%
	\BibitemOpen
	\bibfield  {author} {\bibinfo {author} {\bibfnamefont {B.}~\bibnamefont
			{Bulka}}, \bibinfo {author} {\bibfnamefont {B.}~\bibnamefont {Kramer}},\ and\
		\bibinfo {author} {\bibfnamefont {A.}~\bibnamefont {MacKinnon}},\ }\bibfield
	{title} {\bibinfo {title} {Mobility edge in the three dimensional anderson
			model},\ }\href {https://doi.org/10.1007/bf01312638} {\bibfield  {journal}
		{\bibinfo  {journal} {Zeitschrift f{\"u}r Physik B Condensed Matter}\
		}\textbf {\bibinfo {volume} {60}},\ \bibinfo {pages} {13} (\bibinfo {year}
		{1985})}\BibitemShut {NoStop}%
	\bibitem [{\citenamefont {Laumann}\ \emph {et~al.}(2014)\citenamefont
		{Laumann}, \citenamefont {Pal},\ and\ \citenamefont
		{Scardicchio}}]{Scardicchio_QREM}%
	\BibitemOpen
	\bibfield  {author} {\bibinfo {author} {\bibfnamefont {C.~R.}\ \bibnamefont
			{Laumann}}, \bibinfo {author} {\bibfnamefont {A.}~\bibnamefont {Pal}},\ and\
		\bibinfo {author} {\bibfnamefont {A.}~\bibnamefont {Scardicchio}},\
	}\bibfield  {title} {\bibinfo {title} {Many-body mobility edge in a
			mean-field quantum spin glass},\ }\href
	{https://doi.org/10.1103/PhysRevLett.113.200405} {\bibfield  {journal}
		{\bibinfo  {journal} {Phys. Rev. Lett.}\ }\textbf {\bibinfo {volume} {113}},\
		\bibinfo {pages} {200405} (\bibinfo {year} {2014})}\BibitemShut {NoStop}%
	\bibitem [{\citenamefont {Oganesyan}\ and\ \citenamefont
		{Huse}(2007)}]{Oganesyan2007}%
	\BibitemOpen
	\bibfield  {author} {\bibinfo {author} {\bibfnamefont {V.}~\bibnamefont
			{Oganesyan}}\ and\ \bibinfo {author} {\bibfnamefont {D.~A.}\ \bibnamefont
			{Huse}},\ }\bibfield  {title} {\bibinfo {title} {Localization of interacting
			fermions at high temperature},\ }\href
	{https://doi.org/10.1103/PhysRevB.75.155111} {\bibfield  {journal} {\bibinfo
			{journal} {Phys. Rev. B}\ }\textbf {\bibinfo {volume} {75}},\ \bibinfo
		{pages} {155111} (\bibinfo {year} {2007})}\BibitemShut {NoStop}%
	\bibitem [{\citenamefont {Atas}\ \emph {et~al.}(2013)\citenamefont {Atas},
		\citenamefont {Bogomolny}, \citenamefont {Giraud},\ and\ \citenamefont
		{Roux}}]{Atas1}%
	\BibitemOpen
	\bibfield  {author} {\bibinfo {author} {\bibfnamefont {Y.~Y.}\ \bibnamefont
			{Atas}}, \bibinfo {author} {\bibfnamefont {E.}~\bibnamefont {Bogomolny}},
		\bibinfo {author} {\bibfnamefont {O.}~\bibnamefont {Giraud}},\ and\ \bibinfo
		{author} {\bibfnamefont {G.}~\bibnamefont {Roux}},\ }\bibfield  {title}
	{\bibinfo {title} {Distribution of the ratio of consecutive level spacings in
			random matrix ensembles},\ }\href
	{https://doi.org/10.1103/PhysRevLett.110.084101} {\bibfield  {journal}
		{\bibinfo  {journal} {Phys. Rev. Lett.}\ }\textbf {\bibinfo {volume} {110}},\
		\bibinfo {pages} {084101} (\bibinfo {year} {2013})}\BibitemShut {NoStop}%
	\bibitem [{\citenamefont {Guhr}\ \emph {et~al.}(1998)\citenamefont {Guhr},
		\citenamefont {Müller–Groeling},\ and\ \citenamefont
		{Weidenmüller}}]{Guhr1}%
	\BibitemOpen
	\bibfield  {author} {\bibinfo {author} {\bibfnamefont {T.}~\bibnamefont
			{Guhr}}, \bibinfo {author} {\bibfnamefont {A.}~\bibnamefont
			{Müller–Groeling}},\ and\ \bibinfo {author} {\bibfnamefont {H.~A.}\
			\bibnamefont {Weidenmüller}},\ }\bibfield  {title} {\bibinfo {title}
		{Random-matrix theories in quantum physics: common concepts},\ }\href
	{https://doi.org/https://doi.org/10.1016/S0370-1573(97)00088-4} {\bibfield
		{journal} {\bibinfo  {journal} {Physics Reports}\ }\textbf {\bibinfo {volume}
			{299}},\ \bibinfo {pages} {189 } (\bibinfo {year} {1998})}\BibitemShut
	{NoStop}%
	\bibitem [{\citenamefont {Faleiro}\ \emph {et~al.}(2004)\citenamefont
		{Faleiro}, \citenamefont {G\'omez}, \citenamefont {Molina}, \citenamefont
		{Mu\~noz}, \citenamefont {Rela\~no},\ and\ \citenamefont
		{Retamosa}}]{Faleiro1}%
	\BibitemOpen
	\bibfield  {author} {\bibinfo {author} {\bibfnamefont {E.}~\bibnamefont
			{Faleiro}}, \bibinfo {author} {\bibfnamefont {J.~M.~G.}\ \bibnamefont
			{G\'omez}}, \bibinfo {author} {\bibfnamefont {R.~A.}\ \bibnamefont {Molina}},
		\bibinfo {author} {\bibfnamefont {L.}~\bibnamefont {Mu\~noz}}, \bibinfo
		{author} {\bibfnamefont {A.}~\bibnamefont {Rela\~no}},\ and\ \bibinfo
		{author} {\bibfnamefont {J.}~\bibnamefont {Retamosa}},\ }\bibfield  {title}
	{\bibinfo {title} {Theoretical derivation of $1/f$ noise in quantum chaos},\
	}\href {https://doi.org/10.1103/PhysRevLett.93.244101} {\bibfield  {journal}
		{\bibinfo  {journal} {Phys. Rev. Lett.}\ }\textbf {\bibinfo {volume} {93}},\
		\bibinfo {pages} {244101} (\bibinfo {year} {2004})}\BibitemShut {NoStop}%
	\bibitem [{\citenamefont {Riser}\ \emph {et~al.}(2017)\citenamefont {Riser},
		\citenamefont {Osipov},\ and\ \citenamefont {Kanzieper}}]{Riser1}%
	\BibitemOpen
	\bibfield  {author} {\bibinfo {author} {\bibfnamefont {R.}~\bibnamefont
			{Riser}}, \bibinfo {author} {\bibfnamefont {V.~A.}\ \bibnamefont {Osipov}},\
		and\ \bibinfo {author} {\bibfnamefont {E.}~\bibnamefont {Kanzieper}},\
	}\bibfield  {title} {\bibinfo {title} {Power spectrum of long eigenlevel
			sequences in quantum chaotic systems},\ }\href
	{https://doi.org/10.1103/PhysRevLett.118.204101} {\bibfield  {journal}
		{\bibinfo  {journal} {Phys. Rev. Lett.}\ }\textbf {\bibinfo {volume} {118}},\
		\bibinfo {pages} {204101} (\bibinfo {year} {2017})}\BibitemShut {NoStop}%
	\bibitem [{\citenamefont {Rela\~no}\ \emph {et~al.}(2008)\citenamefont
		{Rela\~no}, \citenamefont {Mu\~noz}, \citenamefont {Retamosa}, \citenamefont
		{Faleiro},\ and\ \citenamefont {Molina}}]{Relano3}%
	\BibitemOpen
	\bibfield  {author} {\bibinfo {author} {\bibfnamefont {A.}~\bibnamefont
			{Rela\~no}}, \bibinfo {author} {\bibfnamefont {L.}~\bibnamefont {Mu\~noz}},
		\bibinfo {author} {\bibfnamefont {J.}~\bibnamefont {Retamosa}}, \bibinfo
		{author} {\bibfnamefont {E.}~\bibnamefont {Faleiro}},\ and\ \bibinfo {author}
		{\bibfnamefont {R.~A.}\ \bibnamefont {Molina}},\ }\bibfield  {title}
	{\bibinfo {title} {Power-spectrum characterization of the continuous gaussian
			ensemble},\ }\href {https://doi.org/10.1103/PhysRevE.77.031103} {\bibfield
		{journal} {\bibinfo  {journal} {Phys. Rev. E}\ }\textbf {\bibinfo {volume}
			{77}},\ \bibinfo {pages} {031103} (\bibinfo {year} {2008})}\BibitemShut
	{NoStop}%
	\bibitem [{\citenamefont {Berkovits}(2020)}]{Berkovits2}%
	\BibitemOpen
	\bibfield  {author} {\bibinfo {author} {\bibfnamefont {R.}~\bibnamefont
			{Berkovits}},\ }\bibfield  {title} {\bibinfo {title} {Super-poissonian
			behavior of the rosenzweig-porter model in the nonergodic extended regime},\
	}\href {https://doi.org/10.1103/PhysRevB.102.165140} {\bibfield  {journal}
		{\bibinfo  {journal} {Phys. Rev. B}\ }\textbf {\bibinfo {volume} {102}},\
		\bibinfo {pages} {165140} (\bibinfo {year} {2020})}\BibitemShut {NoStop}%
	\bibitem [{\citenamefont {Das}\ and\ \citenamefont {Ghosh}(2019)}]{Das1}%
	\BibitemOpen
	\bibfield  {author} {\bibinfo {author} {\bibfnamefont {A.~K.}\ \bibnamefont
			{Das}}\ and\ \bibinfo {author} {\bibfnamefont {A.}~\bibnamefont {Ghosh}},\
	}\bibfield  {title} {\bibinfo {title} {Eigenvalue statistics for generalized
			symmetric and hermitian matrices},\ }\href
	{https://doi.org/10.1088/1751-8121/ab3711} {\bibfield  {journal} {\bibinfo
			{journal} {Journal of Physics A: Mathematical and Theoretical}\ }\textbf
		{\bibinfo {volume} {52}},\ \bibinfo {pages} {395001} (\bibinfo {year}
		{2019})}\BibitemShut {NoStop}%
	\bibitem [{\citenamefont {Altshuler}\ and\ \citenamefont
		{Kravtsov}(2023)}]{Altshuler3}%
	\BibitemOpen
	\bibfield  {author} {\bibinfo {author} {\bibfnamefont {B.~L.}\ \bibnamefont
			{Altshuler}}\ and\ \bibinfo {author} {\bibfnamefont {V.~E.}\ \bibnamefont
			{Kravtsov}},\ }\href {https://doi.org/10.48550/ARXIV.2301.12279} {\bibinfo
		{title} {Random cantor sets and mini-bands in local spectrum of quantum
			systems}} (\bibinfo {year} {2023})\BibitemShut {NoStop}%
	\bibitem [{\citenamefont {Tomasi}\ \emph {et~al.}(2019)\citenamefont {Tomasi},
		\citenamefont {Amini}, \citenamefont {Bera}, \citenamefont {Khaymovich},\
		and\ \citenamefont {Kravtsov}}]{Tomasi1}%
	\BibitemOpen
	\bibfield  {author} {\bibinfo {author} {\bibfnamefont {G.~D.}\ \bibnamefont
			{Tomasi}}, \bibinfo {author} {\bibfnamefont {M.}~\bibnamefont {Amini}},
		\bibinfo {author} {\bibfnamefont {S.}~\bibnamefont {Bera}}, \bibinfo {author}
		{\bibfnamefont {I.~M.}\ \bibnamefont {Khaymovich}},\ and\ \bibinfo {author}
		{\bibfnamefont {V.~E.}\ \bibnamefont {Kravtsov}},\ }\bibfield  {title}
	{\bibinfo {title} {{Survival probability in Generalized Rosenzweig-Porter
				random matrix ensemble}},\ }\href
	{https://doi.org/10.21468/SciPostPhys.6.1.014} {\bibfield  {journal}
		{\bibinfo  {journal} {SciPost Phys.}\ }\textbf {\bibinfo {volume} {6}},\
		\bibinfo {pages} {14} (\bibinfo {year} {2019})}\BibitemShut {NoStop}%
	\bibitem [{\citenamefont {Sanchez-Palencia}\ \emph {et~al.}(2007)\citenamefont
		{Sanchez-Palencia}, \citenamefont {Cl\'ement}, \citenamefont {Lugan},
		\citenamefont {Bouyer}, \citenamefont {Shlyapnikov},\ and\ \citenamefont
		{Aspect}}]{Sanchez1}%
	\BibitemOpen
	\bibfield  {author} {\bibinfo {author} {\bibfnamefont {L.}~\bibnamefont
			{Sanchez-Palencia}}, \bibinfo {author} {\bibfnamefont {D.}~\bibnamefont
			{Cl\'ement}}, \bibinfo {author} {\bibfnamefont {P.}~\bibnamefont {Lugan}},
		\bibinfo {author} {\bibfnamefont {P.}~\bibnamefont {Bouyer}}, \bibinfo
		{author} {\bibfnamefont {G.~V.}\ \bibnamefont {Shlyapnikov}},\ and\ \bibinfo
		{author} {\bibfnamefont {A.}~\bibnamefont {Aspect}},\ }\bibfield  {title}
	{\bibinfo {title} {Anderson localization of expanding bose-einstein
			condensates in random potentials},\ }\href
	{https://doi.org/10.1103/PhysRevLett.98.210401} {\bibfield  {journal}
		{\bibinfo  {journal} {Phys. Rev. Lett.}\ }\textbf {\bibinfo {volume} {98}},\
		\bibinfo {pages} {210401} (\bibinfo {year} {2007})}\BibitemShut {NoStop}%
	\bibitem [{\citenamefont {F{\"u}redi}\ and\ \citenamefont
		{Koml{\'o}s}(1981)}]{Furedi1}%
	\BibitemOpen
	\bibfield  {author} {\bibinfo {author} {\bibfnamefont {Z.}~\bibnamefont
			{F{\"u}redi}}\ and\ \bibinfo {author} {\bibfnamefont {J.}~\bibnamefont
			{Koml{\'o}s}},\ }\bibfield  {title} {\bibinfo {title} {The eigenvalues of
			random symmetric matrices},\ }\href {https://doi.org/10.1007/bf02579329}
	{\bibfield  {journal} {\bibinfo  {journal} {Combinatorica}\ }\textbf
		{\bibinfo {volume} {1}},\ \bibinfo {pages} {233} (\bibinfo {year}
		{1981})}\BibitemShut {NoStop}%
	\bibitem [{\citenamefont {Prosen}\ and\ \citenamefont
		{Robnik}(1993)}]{Prosen1}%
	\BibitemOpen
	\bibfield  {author} {\bibinfo {author} {\bibfnamefont {T.}~\bibnamefont
			{Prosen}}\ and\ \bibinfo {author} {\bibfnamefont {M.}~\bibnamefont
			{Robnik}},\ }\bibfield  {title} {\bibinfo {title} {Energy level statistics
			and localization in sparsed banded random matrix ensemble},\ }\href
	{https://doi.org/10.1088/0305-4470/26/5/029} {\bibfield  {journal} {\bibinfo
			{journal} {Journal of Physics A: Mathematical and General}\ }\textbf
		{\bibinfo {volume} {26}},\ \bibinfo {pages} {1105} (\bibinfo {year}
		{1993})}\BibitemShut {NoStop}%
	\bibitem [{\citenamefont {Albrecht}\ \emph {et~al.}(2009)\citenamefont
		{Albrecht}, \citenamefont {Chan},\ and\ \citenamefont {Edelman}}]{Albrecht1}%
	\BibitemOpen
	\bibfield  {author} {\bibinfo {author} {\bibfnamefont {J.~T.}\ \bibnamefont
			{Albrecht}}, \bibinfo {author} {\bibfnamefont {C.~P.}\ \bibnamefont {Chan}},\
		and\ \bibinfo {author} {\bibfnamefont {A.}~\bibnamefont {Edelman}},\
	}\bibfield  {title} {\bibinfo {title} {Sturm sequences and random eigenvalue
			distributions},\ }\href {https://doi.org/10.1007/s10208-008-9037-x}
	{\bibfield  {journal} {\bibinfo  {journal} {Foundations of Computational
				Mathematics}\ }\textbf {\bibinfo {volume} {9}},\ \bibinfo {pages} {461}
		(\bibinfo {year} {2009})}\BibitemShut {NoStop}%
	\bibitem [{\citenamefont {Allez}\ \emph {et~al.}(2012)\citenamefont {Allez},
		\citenamefont {Bouchaud},\ and\ \citenamefont {Guionnet}}]{Allez2}%
	\BibitemOpen
	\bibfield  {author} {\bibinfo {author} {\bibfnamefont {R.}~\bibnamefont
			{Allez}}, \bibinfo {author} {\bibfnamefont {J.-P.}\ \bibnamefont
			{Bouchaud}},\ and\ \bibinfo {author} {\bibfnamefont {A.}~\bibnamefont
			{Guionnet}},\ }\bibfield  {title} {\bibinfo {title} {Invariant beta ensembles
			and the gauss-wigner crossover},\ }\href
	{https://doi.org/10.1103/PhysRevLett.109.094102} {\bibfield  {journal}
		{\bibinfo  {journal} {Phys. Rev. Lett.}\ }\textbf {\bibinfo {volume} {109}},\
		\bibinfo {pages} {094102} (\bibinfo {year} {2012})}\BibitemShut {NoStop}%
	\bibitem [{\citenamefont {Duy}\ and\ \citenamefont {Nakano}(2016)}]{Duy1}%
	\BibitemOpen
	\bibfield  {author} {\bibinfo {author} {\bibfnamefont {T.~K.}\ \bibnamefont
			{Duy}}\ and\ \bibinfo {author} {\bibfnamefont {F.}~\bibnamefont {Nakano}},\
	}\href {https://doi.org/10.48550/ARXIV.1611.09476} {\bibinfo {title}
		{Gaussian beta ensembles at high temperature: eigenvalue fluctuations and
			bulk statistics}} (\bibinfo {year} {2016})\BibitemShut {NoStop}%
	\bibitem [{\citenamefont {Sorathia}\ \emph {et~al.}(2012)\citenamefont
		{Sorathia}, \citenamefont {Izrailev}, \citenamefont {Zelevinsky},\ and\
		\citenamefont {Celardo}}]{Sorathia1}%
	\BibitemOpen
	\bibfield  {author} {\bibinfo {author} {\bibfnamefont {S.}~\bibnamefont
			{Sorathia}}, \bibinfo {author} {\bibfnamefont {F.~M.}\ \bibnamefont
			{Izrailev}}, \bibinfo {author} {\bibfnamefont {V.~G.}\ \bibnamefont
			{Zelevinsky}},\ and\ \bibinfo {author} {\bibfnamefont {G.~L.}\ \bibnamefont
			{Celardo}},\ }\bibfield  {title} {\bibinfo {title} {From closed to open
			one-dimensional anderson model: Transport versus spectral statistics},\
	}\href {https://doi.org/10.1103/PhysRevE.86.011142} {\bibfield  {journal}
		{\bibinfo  {journal} {Phys. Rev. E}\ }\textbf {\bibinfo {volume} {86}},\
		\bibinfo {pages} {011142} (\bibinfo {year} {2012})}\BibitemShut {NoStop}%
\end{thebibliography}
%
\appendix
\onecolumngrid
\renewcommand\thefigure{\Roman{figure}}  
\setcounter{figure}{0}
\renewcommand\thetable{\Roman{table}}
\setcounter{table}{0}
{
}
\section{Hopping amplitudes}\label{sec_hop}
The $n$th hopping term in \bte\ is denoted by $y_n$ where $\sqrt{2}y_n\sim \chi_{n\beta}$, i.e.~chi distribution with degree of freedom $n\beta = \frac{n}{N^\gamma}$. Then $m$th moment of $y_n$ is
\begin{align}
	\label{eq_hop_moment}
	\mean{y_n^m} = \frac{ \fgamma{\frac{n\beta + m}{2}} }{ \fgamma{\frac{n\beta}{2}} }
\end{align}
Then $y_n\sim \mathcal{N}\del{\sqrt{\dfrac{n\beta}{2}}, \dfrac{1}{4}}$ and $y_n\sim \mathcal{O}(1)$ for $n\beta\gg 1$ and $n\beta\approx 1$, respectively. Contrarily for $n\beta\ll 1$, $z\equiv \log y_n$ has a broad distribution $\sim \exp\del{-z - \dfrac{e^{2z}}{2}}$ with a mean $\mean{z} = \dfrac{1}{2}\del{\log 2 + \fdg{\dfrac{n\beta}{2}}}$, where $\fdg{x}\equiv \dfrac{\Gamma'(x)}{\fgamma{x}}$ is the logarithmic derivative of Gamma function, which admits a Taylor series expansion $\fdg{x} = -\dfrac{1}{x} - \gamma_\text{Euler} + \dfrac{\pi^2}{6}x + \mathcal{O}(x^2)$, where $\gamma_\text{Euler}\approx 0.5772$ is the Euler-Mascheroni constant. Consequently the typical value of $y_n$ is $\exp\del{\mean{\log y_n}}\sim \exp\del{-\dfrac{1}{n\beta}}$ for $n\beta\ll 1$. Hence the typical behavior of the hopping terms can be summarized as
\begin{align}
	\label{eq_hop_typical_1}
	y_n\sim \begin{cases}
		\exp\del{-\dfrac{1}{n\beta}}, & n\beta < 1\Rightarrow n < N^\gamma\\
		\mathcal{O}(1), & n\beta\approx 1\Rightarrow n\approx N^\gamma\\
		\sqrt{n\beta}, & n\beta > 1\Rightarrow n > N^\gamma
	\end{cases}.
\end{align}
However $\log y_n$ has a long-tailed density for $n\beta\ll 1$ where the higher order moments of $y_n$ become important. Using Sterling's approximation $\del{x! \sim \sqrt{2\pi x}\del{\dfrac{x}{e}}^x }$, we get that $\mean{\dfrac{y_n^m}{m!}}^{\frac{1}{m}}\sim \sqrt{\dfrac{e}{2m}}\del{\dfrac{n\beta}{2}}^{\frac{1}{m}}$, which exhibits maximum at $m = 2\log\dfrac{2}{n\beta}$. Then largest possible fluctuation in $y_n$ is
\begin{align*}
	\max\limits_{m}\mean{\dfrac{y_n^m}{m!}}^{\frac{1}{m}}\sim \dfrac{1}{\sqrt{-4\log n\beta}} \sim \dfrac{1}{\sqrt{\gamma\log N}}.
\end{align*}
Thus for $n\beta\ll 1$, $y_n$ can exhibit a value $\sim \dfrac{1}{\sqrt{\gamma\log N}}$ while the typical $y_{n}\sim \exp\del{-\dfrac{1}{n\beta}}$ is much smaller.

\section{Identifying localized states}\label{sec_psi_loc}
\begin{figure}
	\centering
	\includegraphics[width=\textwidth]{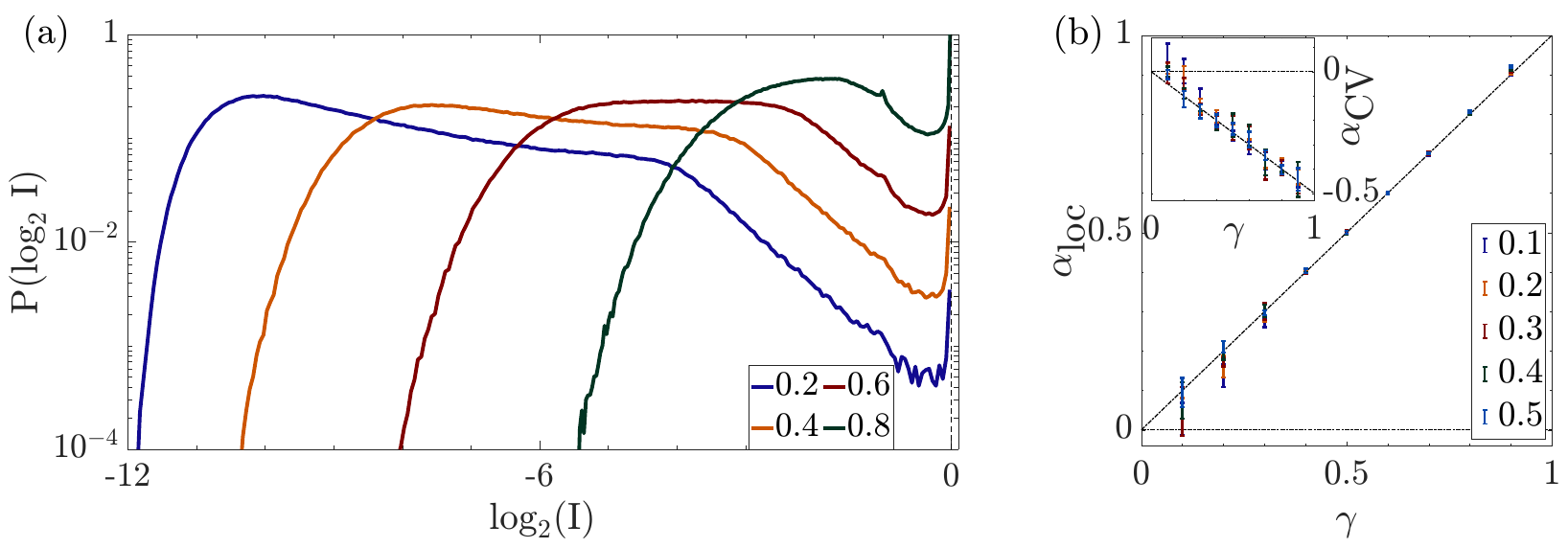} 
	\caption{(a) Density of IPR for various $\gamma$ and system size $N = 8192$.
		(b) system size scaling exponent of $\mean{\nloc}$, the ensemble averaged number of localized states, (i.e. $\mean{\nloc}\propto N^{\aloc}$) as a function of $\gamma$. Inset shows the scaling exponent of the coefficient of variation of $\nloc$, i.e. CV$\del{\nloc}\propto N^{\alpha_\text{CV}}$. The error-bars denote 99.9\% confidence interval obtained from linear fitting in log-log scale.
	}
	\label{fig_n_loc}
\end{figure}
Fig.~\ref{fig_n_loc}(a) shows the density of $\log_2\del{\ipr}$ for \bte\ with $N = 8192$ and various $\gamma$. Here we observe a peak at 0 (i.e.~$\ipr = 1$), indicating the existence of localized states in the NEE phase and we need to find the number of such localized states at a particular $\gamma$. Given a tolerance value $\idel\ll 1$, we consider a state with $\ipr>1-\idel$ to be strongly localized and denote it by $\sloc$. Let, $\nloc$ be the number of $\sloc$ in a single realization of \bte\ according to above definition. We observe that the ensemble averaged $\nloc$ shows a power law behavior w.r.t.~system size, i.e.~$\mean{\nloc}\propto N^{\aloc}$. In Fig.~\ref{fig_n_loc}(b), we show $\aloc$ as a function of $\gamma$ for various $\idel$ and observe that $\aloc\approx \gamma$ for any $\idel$. Next we look at the coefficient of variation (CV) of $\nloc$, CV$\del{\nloc} \equiv \dfrac{ \sqrt{\mean{\nloc^2} - \mean{\nloc}^2} }{\mean{\nloc}}$, which captures the sample-sample fluctuation of the number of $\sloc$. We find that CV$(\nloc)\propto N^{\alpha_\text{CV}}$, where $\alpha_\text{CV}\approx -\dfrac{\gamma}{2}$ for any $\idel$ as shown in the inset of Fig.~\ref{fig_n_loc}(b). Since both $\nloc$ and CV$\del{\nloc}\sim \mathcal{O}(1)$ for $\gamma\to 0$, some finite samples may not have any $\sloc$ in the NEE phase close to $\get$. However deep in the NEE regime, sample-sample fluctuation of $\nloc$ is absent in the thermodynamic limit and we always find $\mathcal{O}(N^\gamma)$ i.e.~an extensive number but a zero fraction of all the eigenstates to be localized.

\section{Energy correlations of localized states}\label{sec_e_loc}
\begin{figure}[b]
	\centering
	\includegraphics[width=\textwidth]{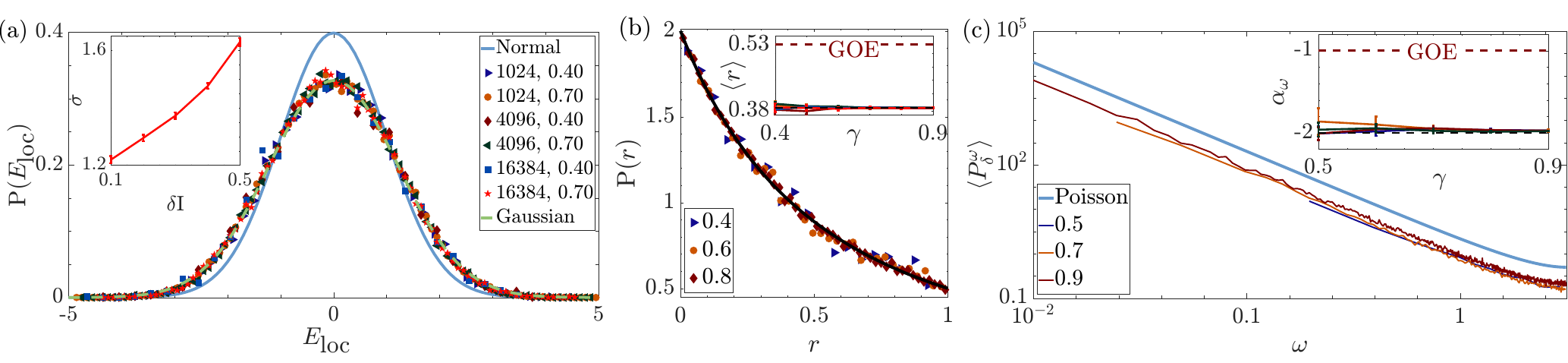} 
	\caption{(a) Density of $\eloc$, energy of localized states for various $N$ and $\gamma$ where $\idel = 0.1$. The dashed line denotes Gaussian fit with mean 0 and standard deviation, $\sigma = 1.219\pm 0.013$ while solid line denotes normal distribution, $\mathcal{N}(0, 1)$. Inset shows $\sigma$ vs. $\idel$.
		(b) density of $r$ of $\eloc$ for various $\gamma$ where $N = 8192$ and $\idel = 0.3$. Solid line indicate the density of $r$ for Poisson ensemble \cite{Atas1}. The inset shows ensemble average of $r$ vs. $\gamma$ for $N = 2048, 8192, 32768$ and $\idel = 0.2, 0.4$.
		(c) Power spectrum of noise for various $\gamma$ where $N = 8192$ and $\idel = 0.3$. Solid line indicate the Power spectrum for Poisson ensemble \cite{Faleiro1}. The inset shows $\alpha_\omega$, the slope of the power spectrum $\del{ \mean{P_\omega^\delta}\propto \omega^{\alpha_\omega} }$ vs. $\gamma$ for $N = 8192, 16384, 32768$ and $\idel = 0.3, 0.5$. Error-bars denote 95\% confidence interval.
	}
	\label{fig_e_loc}
\end{figure}
Let us denote the energy levels of the localized states by $\eloc$. In Fig.~\ref{fig_e_loc}(a), we show the density of $\eloc$ for various $\gamma$ where $N = 8192$ and $\idel = 0.1$. We observe that for a fixed $\idel$, the density of $\eloc$ has a Gaussian shape independent of both $N$ and $\gamma$, which becomes sharper as we decrease $\idel$. In the inset of Fig.~\ref{fig_e_loc}(a), we show $\sigma$, the standard deviation of the Gaussian fit as a function of $\idel$ and we see that $\sigma\to 1$ as $\idel\to 0$. Thus for $\sloc$ with $\ipr\to 1$, the Density of States (DOS) follows a normal distribution. However the global shape of the DOS does not reflect the nature of correlations present among $\eloc$'s. The degree of correlation present in a local energy window is captured by $r$, the ratio of level spacing \cite{Atas1}.
In Fig.~\ref{fig_e_loc}(b), we show the density of $r$ for various $\gamma$ where $N = 8192$ and $\idel = 0.3$. We observe that all the densities match nicely with $\text{P}_\text{Poisson}(r) = \dfrac{2}{(1+r)^2}$. In the inset of Fig.~\ref{fig_e_loc}(b), we show the ensemble averaged $r$ as a function of $\gamma$ for various $N$ and $\idel$. We observe that irrespective of $N, \gamma$ or $\idel$, $\mean{r}\approx 0.38$ which is expected of a Poisson ensemble. Thus short-range correlation is absent in the energy levels of localized states.

Complimentary to the short-range measures like ratio of level spacing, the long-range spectral correlations are captured by $P_\omega^\delta$, the power-spectrum (squared modulus of Fourier transform) of $\delta_n$ statistics.
As $\eloc/\sigma$ follows a normal distribution, the unfolding becomes trivial where the unfolded energy $\cale_\text{loc} = \dfrac{\nloc}{2}\del{ 1 + \ferf{\dfrac{\eloc}{\sqrt{2}}} }$. The ensemble averaged power spectrum of noise in such unfolded energy levels exhibits a power-law behavior w.r.t.~$\omega$ as shown in Fig.~\ref{fig_e_loc}(c), i.e.~$\mean{P_\omega^\delta}\propto \omega^{\alpha_\omega}$. The inset of Fig.~\ref{fig_e_loc}(c) shows that irrespective of $N, \gamma$ or $\idel$, the exponent $\alpha_\omega\approx -2$ similar to Poisson ensemble. Thus we can conclude that the energy levels of localized states are uncorrelated at any distance.

\begin{figure}[t]
	\centering
	\includegraphics[width=\textwidth]{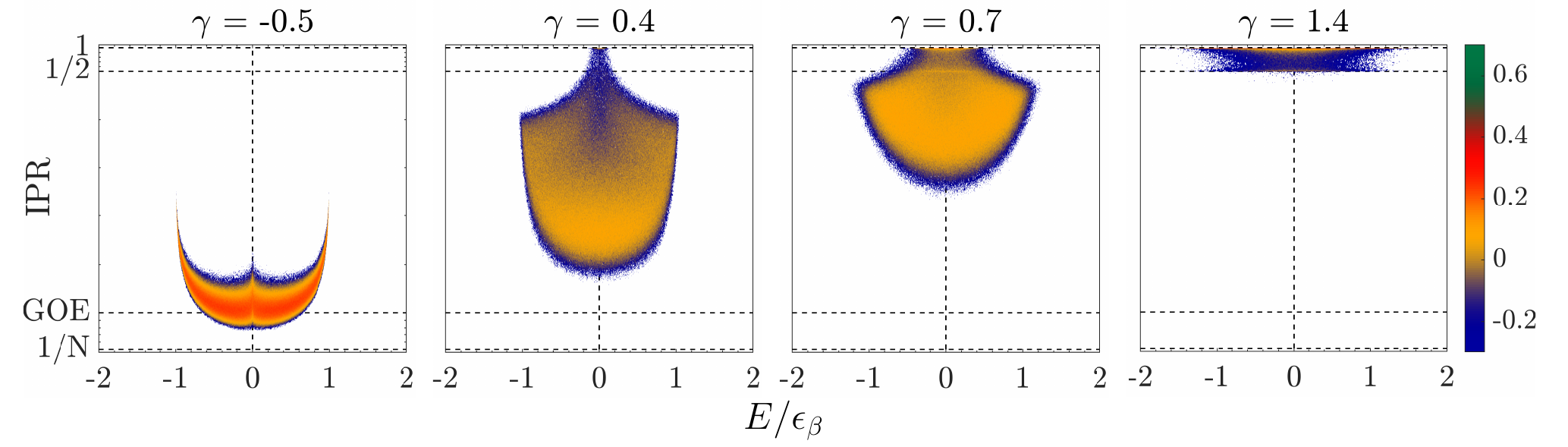} 
	\caption{Joint density of IPR and energy for various $\gamma$ and $N = 8192$ obtained from 128 disordered samples. The colorbar indicates values of $\prob{\ipr, \dfrac{E}{\epsilon_\beta}}$ in $\log_N$ scale where $\epsilon_\beta\equiv \sqrt{4 + 2N^{1-\gamma}}$ is the global bandwidth.
	}
	\label{fig_I_e_PD}
\end{figure}
\begin{figure}[h]
	\centering
	\includegraphics[width=0.6\textwidth]{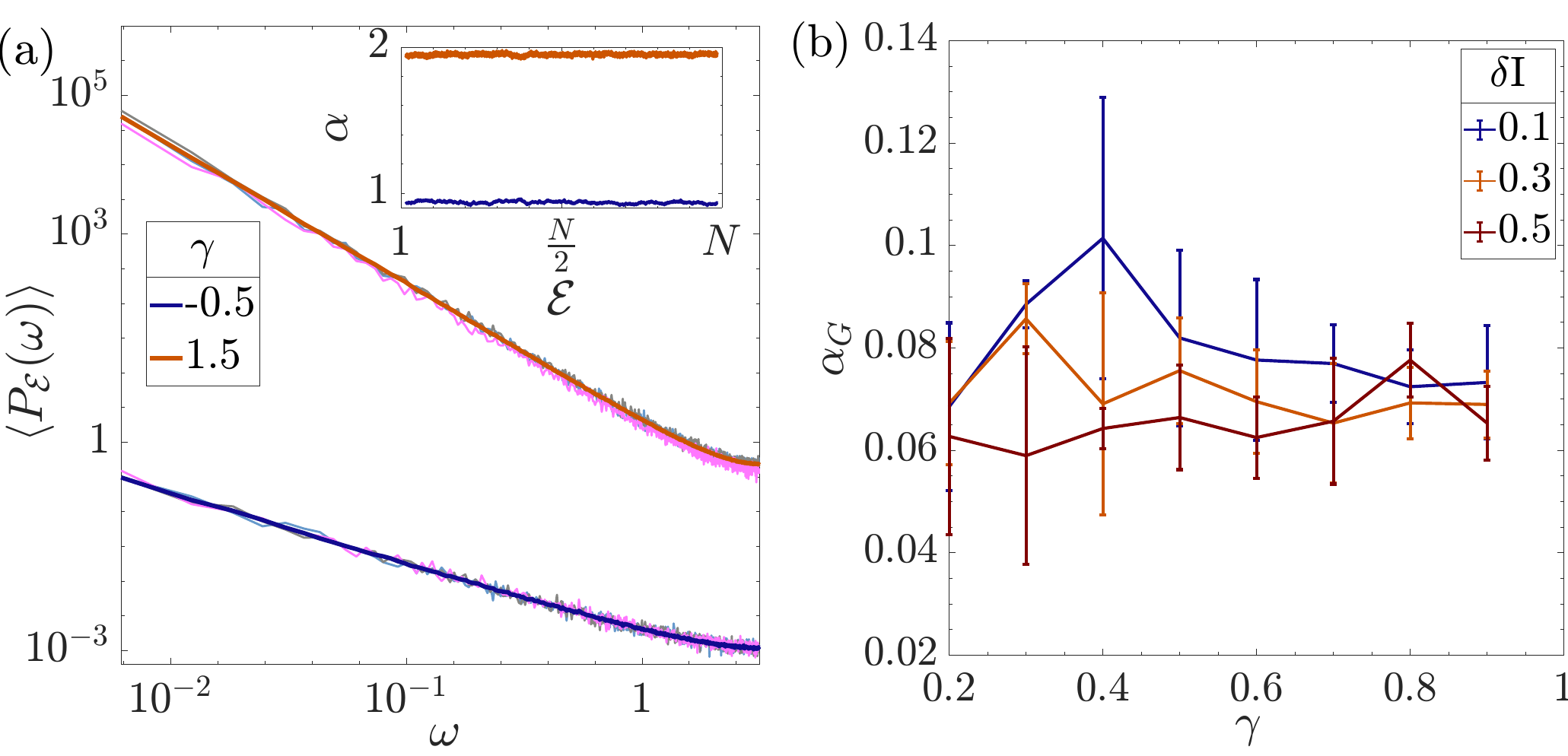} 
	\caption{(a) ensemble averaged power spectrum of noise vs.~dimensionless frequency, $\omega$ for various $\gamma$ where $\Delta\cale = 1024$ and $N = 32768$. For each $\gamma$, solid lines denote $\mean{P_\cale(\omega)}$ at three randomly chosen $\cale$ and the dashed curves denote $\mean{P_\cale(\omega)}$ averaged over $\cale$. Inset shows exponent $\alpha$ vs.~unfolded energy where $\mean{P_\cale(\omega)}\propto \omega^{-\alpha}$.
	(b) system size scaling of $\mean{\eg} \propto N^{\alpha_G}$ for various $\delta\ipr$ where $\del{-\eg, \eg}$ is the energy bound of localized state (has $\ipr>1-\delta\ipr$).
	}
	\label{fig_sup_4}
\end{figure}

In Fig.~\ref{fig_I_e_PD}, we show the joint density of IPR and energy for various values of $\gamma$. Such plots show the coexistence localized states in the NEE phase as well as helps us identify $(-\eg, \eg)$, the energy bound of localized states. Fig.~\ref{fig_sup_4}(b) shows that the ensemble average $\mean{\eg}$ does not scale with system size for any value of $\gamma$ irrespective of $\idel$.

In Fig.~\ref{fig_sup_4}(a), we show that the long-range energy correlations are homogeneous over the energy spectrum in the ergodic and localized phase.

\section{Exponential decay of eigenstates}\label{sec_exp}
\begin{figure}
	\centering
	\includegraphics[width=0.3\textwidth]{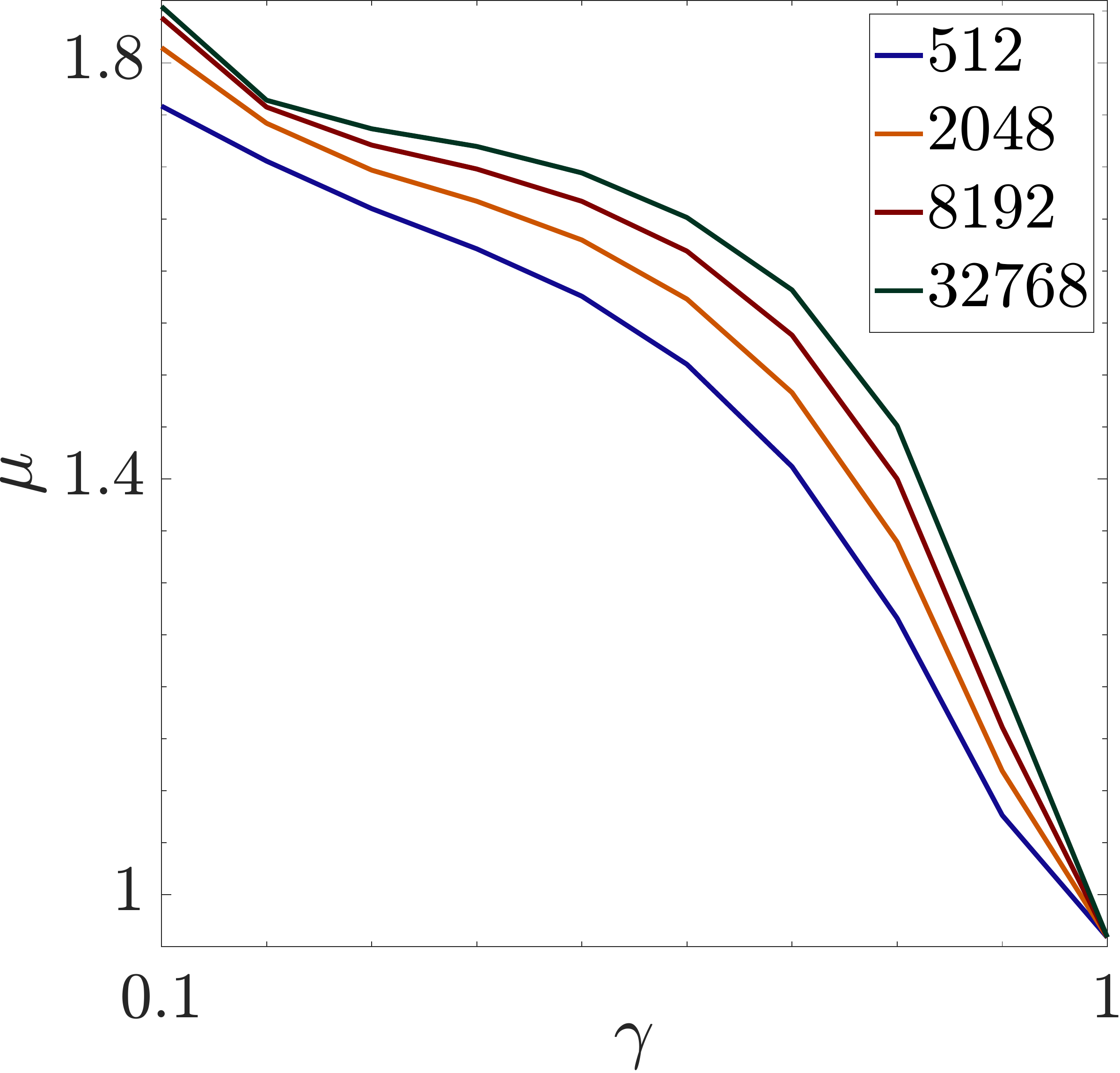}
	\caption[]{Metric $\mu$ from Eq.~\eqref{eq_mu_exp_decay} vs.~$\gamma$ for various $N$.
	}
	\label{fig_exp_decay}
\end{figure}

The equivalence to 1D Anderson model in a local spatial window indicates that the eigenstates of \bte\ in the NEE phase are exponentially decaying. To verify this we look at the following metric
\begin{align}
	\label{eq_mu_exp_decay}
	\mu = \frac{8\xi_g}{\xi_\shn},\quad \begin{matrix}
		\xi_g &=& \sqrt{\sum_j j^2\abs{\Psi(j)}^2 - \del{\sum_j j\abs{\Psi(j)}^2}^2}\\
		\xi_\shn &=& 2.07\exp\del{\shn}
	\end{matrix}
\end{align}
where $\xi_g$ and $\xi_\shn$ are geometric and entropic localization lengths, respectively. For exponentially localized states, $\xi_\shn\approx 8\xi_g$, i.e. $\mu\approx 1$. Contrarily for sparse localized states (where more than one dominant components exist in a single localization length), $\xi_\shn\ll 8\xi_g$, i.e. $\mu\gg 1$. In Fig.~\ref{fig_exp_decay}, we show the median of $\mu$ as a function of $\gamma$ for various $N$. Since $\mu\sim \mathcal{O}(1)$ irrespective of $\gamma$ and $N$, the eigenstates of \bte\ are indeed exponentially decaying.
\end{document}